\documentstyle[preprint,aps]{revtex}
\begin{document}
\draft
\title{On the Low-energy Effective Action of $N=2$
Supersymmetric Yang-Mills Theory }
\author{M. Chaichian${}^{a,b}$, 
W. F. Chen${}^{a,b}$ \thanks{ICSC-World Laboratory, Switzerland}
 and C. Montonen${}^{b}$\thanks{On leave from Theoretical Physics Division, 
Department of Physics, University of Helsinki} }
\address{${}^a$ High Energy Physics Division, Department of Physics, 
University of Helsinki\\
${}^b$ Helsinki Institute of Physics, 
P.O. Box 9 (Siltavuorenpenger 20 C)\\
FIN-00014 University of Helsinki, Finland }

\maketitle

\begin{abstract}
We investigate the perturbative part of Seiberg's low-energy effective 
action of $N=2$ supersymmetric Yang-Mills theory in Wess-Zumino gauge 
in the conventional effective field theory technique. Using the method 
of constant field approximation and restricting the effective action 
with at most two derivatives and not more than four-fermion couplings, 
we show some features of the low-energy effective action given by Seiberg 
based on $U(1)_R$ anomaly and non-perturbative $\beta$-function arguments.
\end{abstract}
 
\vspace{3ex}

\begin{flushleft}
{\bf 1.~ Introduction}
\end{flushleft}

One cannot but be impressed by the steady increase in our knowledge of 
the dynamics of supersymmetric gauge theories ever since their invention. 
The rate of progress has been very rapid in recent years, following the 
seminal contribution by Seiberg and Witten$\cite{sw}$, combining the ideas 
of holomorphicity$\cite{sei}$ and duality$\cite{mo}$.

The web of arguments leading to the explicit results consists of a skilful
combination of perturbative and nonperturbative arguments, formal
considerations and physical reasoning. It should be checked by explicit
computations, whenever possible, that no unexpected failure of these 
arguments occurs. This paper is a modest contribution in that direction.

The subject of our study is that well-studied object, namely the low-energy
effective action of an $N=2$ super-Yang-Mills theory with the gauge group 
$SU(2)$:
\begin{eqnarray}
\Gamma =\frac{1}{16\pi}\mbox{Im}{\int}d^4xd^2{\theta}d^2\tilde{\theta}
\left[ \frac{1}{2}{\tau}{\Psi}^2
+\frac{i}{2\pi}{\Psi}^2\,\ln\frac{{\Psi}^2}{{\Lambda}^2}
+\sum_{n=1}^{\infty}A_n
\left(\frac{{\Lambda}^2}{{\Psi}^2}\right)^{2n}{\Psi}^2\right] \, ,
\label{eq1}
\end{eqnarray}
where $\displaystyle \tau=\frac{\theta}{2\pi}+\frac{4\pi i}{g^2}$
is the modular parameter and ${\Psi}$ the $N=2$ superfield describing
the light-field degrees of freedom. The logarithmic term represents 
the one-loop perturbative result and was first obtained by Di Vecchia 
et al.${\cite{dmnp}}$ in a calculation where they coupled the gauge 
superfield to an $N=2$ matter supermultiplet and integrated out the 
latter. Subsequently, Seiberg$\cite{sei}$ used the anomalous transformation 
behaviour under $U(1)_R$ and holomorphicity to argue that the full 
low-energy effective action should take the form ({\ref{eq1}), where 
the infinite series arises from nonperturbative contributions (instantons).
The Seiberg-Witten solution $\cite{sw}$ gives the explicit form of this 
part of $\Gamma$.

The form ({\ref{eq1}) has been confirmed by calculations in $N=1$ superspace
and in harmonic superspace, extending the result to nonleading terms in the
number of derivatives $\cite{dgr,pw,ov,ma,ke}$. Independent confirmation has
been obtained from $M$-theory $\cite{oo}$. 

In this paper we set out to check the perturbative part of the effective 
action by a very down-to-earth, conventional calculation. In the Higgs 
phase of the theory, the $SU(2)$ gauge symmetry breaks down to $U(1)$, 
and the super-Higgs mechanism splits the supermultiplet into a massive 
one and a massless one. The effective action of the massless fields 
should be obtained by integrating out the heavy fields. Thus our approach 
is very close in spirit to ${\cite{dmnp}}$, although the actual
computations are different. 

Even this modest programme we cannot carry out fully. We report here the 
computation of heavy fermion determinant. Reassuringly, we find that the form
(\ref{eq1}) is reproduced. Although no unexpected surprises were unearthed
by our calculation, we still hope that it has some pedagogical value in
showing explicitly how the effective action arises. 

In section 2, we describe the model and exhibit the Higgs mechanism. Section 3
contains the computation of the heavy fermion determinant using the constant
field approximation. The detailed calculations of the fermion eigenvalues and
their degeneracies, which contain some subtle points, are given in Appendix B. 
In section 4 we present a discussion of the results. In the pedagogical
vein of this paper, we give in Appendix A the component form of the low-energy
effective action.

\begin{flushleft}
{\bf 2.~Super-Higgs Mechanism and Splitting of $N=2$ Supermultiplet }
\end{flushleft}

The classical action of 
$N=2$ supersymmetric Yang-Mills theory with gauge group
$SU(2)$ reads as follows$\cite{dhd}$, 
\begin{eqnarray}
S&=& {\int}d^4 x\left[-\frac{1}{4}G_{\mu\nu}^aG^{\mu\nu a}
+D_{\mu}{\varphi}^{\dagger a}D^{\mu}{\varphi}^a
+i\bar{\psi}^a{\gamma}^{\mu}D_{\mu}^{ab}{\psi}^b\right.\nonumber\\[2mm]
&+&\left. \frac{ig}{\sqrt{2}}{\epsilon}^{abc}\bar{\psi}^c
[(1-{\gamma}_5){\varphi}^a
+(1+{\gamma}_5){\varphi}^{\dagger a}]{\psi}^b
+\frac{g^2}{2}{\epsilon}^{abc}{\epsilon}^{ade}
{\varphi}^b{\varphi}^{\dagger c}{\varphi}^{d}{\varphi}^{\dagger e}\right]\,,
\label{eq2} 
\end{eqnarray}
where
\begin{eqnarray}
G_{\mu\nu}^a&=&{\partial}_{\mu}K_{\nu}^a-{\partial}_{\nu}K_{\mu}^a
-g{\epsilon}^{abc}K_{\mu}^bK_{\nu}^c,~~
 D_{\mu}{\varphi}^a={\partial}_{\mu}{\varphi}^a
-g {\epsilon}^{abc}K_{\mu}^b{\varphi}^c,
\nonumber\\[2mm]
{\varphi}^a&=&\frac{1}{\sqrt{2}}(S^a+iP^a),~~ 
{\varphi}^{\dagger a}=\frac{1}{\sqrt{2}}(S^a-iP^a), ~~a=1,2,3\,.\nonumber
\end{eqnarray}
 The bosonic part of the action (\ref{eq2}) 
is similar to the Georgi-Glashow model
in the Bogomol'nyi-Prasad-Sommerfield (BPS) limit. 
In addition to the fermionic term and Yukawa interaction term, this action
 has the scalar potential
\begin{eqnarray}
V(\varphi)=-\frac{g^2}{2}{\epsilon}^{abc}{\epsilon}^{ade}
{\varphi}^b{\varphi}^{\dagger c}
{\varphi}^d{\varphi}^{\dagger e}
{\equiv}g^2\mbox{Tr}\left([\varphi,{\varphi}^{\dagger}]\right)^2\,.
\end{eqnarray}
The unbroken supersymmetry requires that in the ground state the
scalar potential must vanish, i.e.
\begin{eqnarray}
[\varphi,{\varphi}^{\dagger}]&=&0\,.
\label{eq:h2}
\end{eqnarray}
(\ref{eq:h2}) means that ${\varphi}^{\dagger}$ and ${\varphi}$ should
commute. Owing to the gauge freedom, one can choose$\cite{dhd}$  
\begin{eqnarray}
\langle S^a \rangle =v{\delta}^{a3},~~\langle P^a \rangle =0 \, ,
\label{eq:c1}
\end{eqnarray}
where $v$ is a real constant.   
For $v{\neq}0$ the theory is in the Higgs phase
and exhibits gauge symmetry breaking.
In an unitary gauge we have
\begin{eqnarray}
S^{T}=\left(0,0, S+v \right)  \, ,
\label{ug} 
\end{eqnarray}
and the classical Lagrangian can be written as follows,
\begin{eqnarray}
{\cal L}={\cal L}_V+{\cal L}_S+{\cal L}_P+{\cal L}_F+{\cal L}_Y\, ,
\end{eqnarray}
where ${\cal L}_V$, ${\cal L}_S$, ${\cal L}_P$, ${\cal L}_F$ and ${\cal L}_Y$
denote respectively the vector field, the scalar field,
the scalar potential, 
the fermionic and the Yukawa part, 
\begin{eqnarray}
{\cal L}_V&=&  -\frac{1}{4}({\partial}^{\mu}A^{\nu}
-{\partial}^{\nu}A^{\mu}) ({\partial}_{\mu}A_{\nu}-{\partial}_{\nu}A_{\mu}) 
-\frac{1}{2}({\partial}_{\mu}W_{\nu}^+-{\partial}_{\nu}W_{\mu}^+) 
({\partial}^{\mu}W^{-\nu}-{\partial}^{\nu}W^{-\mu})\nonumber  \\[2mm]
&-&i g[({\partial}^{\mu}W_{\nu}^+W_{\mu}^-
-{\partial}^{\mu}W_{\nu}^-W_{\mu}^+)A^{\nu}+
({\partial}_{\nu}W_{\mu}^-W^{+\mu}
-{\partial}_{\nu}W^+_{\mu}W^{\mu -})A^{\nu}\nonumber\\[2mm]
&+&({\partial}^{\mu}A^{\nu}
-{\partial}^{\nu}A^{\mu})W_{\mu}^+W_{\nu}^-]
+g^2(-W_{\mu}^+W^{-\mu}A_{\nu}A^{\nu}+W_{\mu}^+W_{\nu}^-A^{\mu}A^{\nu})
\nonumber\\
&+& \frac{g^2}{2}W^{+\mu}W^{-\nu}(W^+_{\mu}W^-_{\nu}-W^-_{\mu} W^+_{\nu});
\label{eq8}
\end{eqnarray}
\begin{eqnarray}
{\cal L}_S&=&\frac{1}{2}\partial_{\mu}P\partial^{\mu}P
+\frac{1}{2}\partial_{\mu}S\partial^{\mu}S+\partial_{\mu}P^+\partial_{\mu}P^-
+igA^{\mu}(\partial_{\mu}P^-P^+-\partial_{\mu}P^+P^-)\nonumber\\
&+&igP(\partial^{\mu}P^+W_{\mu}^--\partial^{\mu}P^-W_{\mu}^+)+
ig\partial^{\mu}P(W_{\mu}^+P^--W_{\mu}^-P^+)+g^2P^2W^{+\mu}W^-_{\mu}\nonumber\\
&+&g^2(S+v)^2W^{+\mu}W^-_{\mu}+g^2A^{\nu}A_{\nu}P^{+}P^--g^2(W_{\mu}^+P^--
W_{\mu}^-P^+)A^{\mu}P\nonumber\\
&-&\frac{g^2}{2}(W^{\mu +}P^--W^{\mu -}P^+)^2.
\end{eqnarray}
\begin{eqnarray}
{\cal L}_P=g^2(S+v)^2P^+P^-,
\end{eqnarray}
where
\begin{eqnarray}
&& W_{\mu}^+{\equiv}\frac{1}{\sqrt{2}}(K_{\mu}^{1}
-iK_{\mu}^{2})~,~
W_{\mu}^-{\equiv}\frac{1}{\sqrt{2}}(K_{\mu}^{1}
+iK_{\mu}^{2})~,~K_{\mu}^3 {\equiv}A_{\mu} \,;\nonumber\\
&&P^+ {\equiv}\frac{1}{\sqrt{2}}(P^1-iP^2)~,~P^-{\equiv}
 \frac{1}{\sqrt{2}}(P^1+iP^2)~,~P^3{\equiv}P.
\end{eqnarray}
The above Lagrangians  clearly show that $W_{\mu}^{\pm}$ and $P^{\pm}$
become massive with mass $m{\equiv}|gv|$ while $A_{\mu}$, $S$ and $P$
remain massless.

Up to some total derivative term, the bosonic part of the Lagrangian 
can be written as following form,
\begin{eqnarray} 
{\cal L}_B&=&{\cal L}_V+{\cal L}_S+{\cal L}_P\nonumber\\
&=&-\frac{1}{4}F_{\mu\nu}F^{\mu\nu}+\frac{1}{2}\partial_{\mu}P\partial^{\mu}P
+\frac{1}{2}\partial_{\mu}S\partial^{\mu}S+\frac{1}{2}
W^{+\mu}\left[
{\eta}_{\mu\nu}D^{\dagger\alpha}D_{\alpha}-D_{\nu}^{\dagger}D_{\mu}
-igF_{\mu\nu}\right]W^{-\nu}\nonumber\\[2mm]
&+&\frac{1}{2}W^{-\mu}\left[
{\eta}_{\mu\nu}D^{\alpha}D^{\dagger}_{\alpha}
-D_{\nu}D_{\mu}^{\dagger}+igF_{\mu\nu}\right]W^{+\nu}
+g^2[P^2+(S+v)^2]W_{\mu}^+W^{\mu -}\nonumber\\
&+&\frac{1}{2}P^+(-\partial^{\mu}\partial_{\mu}
+2igA_{\mu}\partial^{\mu}+g^2A_{\mu}A^{\mu})P^-
+\frac{1}{2}P^-(-\partial^{\mu}\partial_{\mu}
-2igA_{\mu}\partial^{\mu}+g^2A_{\mu}A^{\mu})P^+
\nonumber\\
&+&\frac{1}{2}W_{\mu}^+(-igP\partial^{\mu}+ig\partial^{\mu}P-g^2A_{\mu}P)P^-
+\frac{1}{2}P^-(igP\partial^{\mu}+2ig\partial^{\mu}P-g^2A_{\mu}P)W_{\mu}^+
\nonumber\\
&+&\frac{1}{2}P^+(-2ig\partial^{\mu}P-igP\partial^{\mu}-g^2A^{\mu}P)W_{\mu}^-
+\frac{1}{2}W_{\mu}^-(-ig\partial^{\mu}P+igP\partial^{\mu}-g^2A^{\mu}P)P^+
\nonumber\\
&+& \frac{g^2}{2}W^{+\mu}W^{-\nu}(W^+_{\mu}W^-_{\nu}-W^-_{\mu} W^+_{\nu})
-\frac{g^2}{2}(W^+_{\mu}P^--W^-_{\mu}P^+)^2\nonumber\\ 
&=&-\frac{1}{4}F_{\mu\nu}F^{\mu\nu}
+\partial_{\mu}{\phi}^{*}\partial^{\mu}{\phi}
+\frac{1}{2}W^{+\mu}{\Delta}_{\mu\nu}W^{-\nu}
+\frac{1}{2}W^{-\mu}{\Delta}^{\dagger}_{\mu\nu}W^{+\nu}
+\frac{1}{2}P^+\Delta P^-\nonumber\\
&+&\frac{1}{2}P^-\Delta^{\dagger}P^+
+\frac{1}{2}W^{+\mu}\Delta_{\mu}P^-+\frac{1}{2}P^-\tilde{\Delta}_{\mu}W^{+\mu}
+\frac{1}{2}P^+\tilde{\Delta}_{\mu}^{\dagger}W^{-\mu}
+\frac{1}{2}W^{-\mu}\Delta_{\mu}^{\dagger}P^+\nonumber\\
&+& \frac{g^2}{2}W^{+\mu}W^{-\nu}(W^+_{\mu}W^-_{\nu}-W^-_{\mu} W^+_{\nu})
-\frac{g^2}{2}(W^+_{\mu}P^--W^-_{\mu}P^+)^2 \, ,
\end{eqnarray} 
where 
\begin{eqnarray} 
{\Delta}_{\mu\nu}
&{\equiv}&{\eta}_{\mu\nu}D^{\dagger\alpha}D_{\alpha}-D_{\nu}^{\dagger}D_{\mu}
-igF_{\mu\nu}+g^2|\sqrt{2}\phi+v|^2{\eta}_{\mu\nu},\nonumber\\[2mm]
{\Delta}_{\mu\nu}^{\dagger}
&=&{\eta}_{\mu\nu}D^{\alpha}D_{\alpha}^{\dagger}
-D_{\nu}D_{\mu}^{\dagger}
+igF_{\mu\nu}+g^2|\sqrt{2}\phi+v|^2{\eta}_{\mu\nu},\nonumber\\[2mm]
\Delta_{\mu}&{\equiv}&-igP\partial^{\mu}+ig\partial^{\mu}P-g^2A_{\mu}P,~
\tilde{\Delta}_{\mu}{\equiv}igP\partial^{\mu}+2ig\partial^{\mu}P-g^2A_{\mu}P,
\nonumber\\
\Delta_{\mu}^{\dagger}&=&-ig\partial^{\mu}P+igP\partial^{\mu}-g^2A^{\mu}P,
~\tilde{\Delta}_{\mu}^{\dagger}
=-2ig\partial^{\mu}P-igP\partial^{\mu}-g^2A^{\mu}P,\nonumber\\
\Delta &=& -\partial^{\mu}\partial_{\mu}
+2igA_{\mu}\partial^{\mu}+g^2A_{\mu}A^{\mu},~
\Delta^{\dagger}=-\partial^{\mu}\partial_{\mu}
-2igA_{\mu}\partial^{\mu}+g^2A_{\mu}A^{\mu}, \nonumber\\
D_{\mu}&=&\partial_{\mu}-igA_{\mu}, ~D_{\mu}^{\dagger}
=\partial_{\mu}+igA_{\mu}, ~\phi{\equiv}\frac{1}{\sqrt{2}}(S+iP).	
\end{eqnarray}

To show that the spinor fields split massive and massless parts explicitly, 
we make some operations 
on ${\cal L}_F$ and ${\cal L}_Y$. The spinor part is
\begin{eqnarray}
{\cal L}_F&=&i\bar{\psi}^1{\gamma}^{\mu}{\partial}_{\mu}{\psi}^1+
i\bar{\psi}^2{\gamma}^{\mu}{\partial}_{\mu}{\psi}^2
+i\bar{\psi}^3{\gamma}^{\mu}{\partial}_{\mu}{\psi}^3\nonumber\\[2mm]
&+& \frac{g}{\sqrt{2}}\bar{\psi}^1(W_{\mu}^+-W_{\mu}^-)
{\gamma}^{\mu}{\psi}^3+ig \bar{\psi}^1A_{\mu}{\gamma}^{\mu}{\psi}^2
\nonumber\\[2mm]
&+&ig \bar{\psi}^1A_{\mu}{\gamma}^{\mu}{\psi}^2+
 \frac{ig}{\sqrt{2}}\bar{\psi}^2
{\gamma}^{\mu}(W_{\mu}^++W_{\mu}^-){\psi}^3
\nonumber\\[2mm]
&-&\frac{ig}{\sqrt{2}}\bar{\psi}^3{\gamma}^{\mu}
(W_{\mu}^++W_{\mu}^-){\psi}^2-
\frac{g}{\sqrt{2}}\bar{\psi}^3{\gamma}^{\mu}(W_{\mu}^+-W_{\mu}^-){\psi}^1 \, .
\label{fp}
\end{eqnarray}
As for the Yukawa part, we first write it in chiral spinors,
\begin{eqnarray}
{\cal L}_Y=i\sqrt{2}gf^{abc}\bar{\psi}_L^c{\varphi}^a{\psi}_R^b+i
\sqrt{2}gf^{abc}\bar{\psi}_R^c{\varphi}^{\dagger  a}{\psi}_L^b \, ,
\label{yu}
\end{eqnarray}
where ${\psi}_L=\displaystyle \frac{1}{2}(1+{\gamma}_5){\psi}$ and 
${\psi}_R=\displaystyle \frac{1}{2}(1-{\gamma}_5){\psi}$.
In the unitary gauge (\ref{yu}) becomes
\begin{eqnarray}
{\cal L}_Y&=&ig(\sqrt{2}{\phi}+v)
(\bar{\psi}_L^2{\psi}_R^1-\bar{\psi}_L^1{\psi}_R^2)
+ig (\sqrt{2}{\phi}^*+v) (\bar{\psi}_R^2{\psi}_L^1
-\bar{\psi}_R^1{\psi}_L^2)\nonumber\\
&+&\frac{g}{\sqrt{2}}(P^++P^-)\left[(\bar{\psi}_R^3{\psi}_L^2
-\bar{\psi}_R^2{\psi}_L^3)-(\bar{\psi}_L^3{\psi}_R^2
-\bar{\psi}_L^2{\psi}_R^3)\right]\nonumber\\
&+&\frac{ig}{\sqrt{2}}(P^+-P^-)\left[(\bar{\psi}_R^1{\psi}_L^3
-\bar{\psi}_R^3{\psi}_L^1)-(\bar{\psi}_L^1{\psi}_R^3
-\bar{\psi}_L^3{\psi}_R^1)\right] \, .
\end{eqnarray}
Denoting
\begin{eqnarray}
{\Psi}_1{\equiv} \frac{1}{\sqrt{2}}({\psi}^1+i{\psi}^2)\,,\,
   {\Psi}_2{\equiv} \frac{1}{\sqrt{2}}({\psi}^1-i{\psi}^2)\, ,\,
{\Psi}{\equiv}{\psi}^3 \, , 
\end{eqnarray}
we write ${\cal L}_F$ and ${\cal L}_Y$ in terms of these new fields, 
\begin{eqnarray}
{\cal L}_Y&=&
-g\bar{\Psi}_1\left[\frac{1}{\sqrt{2}}(1-\gamma_5)\phi
+\frac{1}{\sqrt{2}}(1+\gamma_5)\phi^*+v\right]{\Psi}_1\nonumber\\
&+&g\bar{\Psi}_2 \left[\frac{1}{\sqrt{2}}(1-\gamma_5)\phi
+\frac{1}{\sqrt{2}}(1+\gamma_5)\phi^*+v\right]{\Psi}_2
\nonumber\\
&-& igP^+\bar{\Psi}\gamma_5\Psi_1+igP^-\bar{\Psi}\gamma_5\Psi_2
-ig\bar{\Psi}_1 \gamma_5\Psi P^-+ig\bar{\Psi}_2\gamma_5\Psi P^+ \, ,
\\[2mm]
{\cal L}_F&=&i\bar{\Psi}_1{\gamma}^{\mu}{\partial}_{\mu}{\Psi}_1+
i\bar{\Psi}_2{\gamma}^{\mu}{\partial}_{\mu}{\Psi}_2
+i\bar{\Psi}{\gamma}^{\mu}{\partial}_{\mu}{\Psi}\nonumber\\[2mm]
&+&g\bar{\Psi}_1{\gamma}^{\mu}A_{\mu}{\Psi}_1-
g\bar{\Psi}_2{\gamma}^{\mu}A_{\mu}{\Psi}_2 \nonumber\\[2mm]
&+&g\bar{\Psi}_2{\gamma}^{\mu}W_{\mu}^+{\Psi}-
g\bar{\Psi}_1{\gamma}^{\mu}W_{\mu}^-{\Psi}\nonumber\\[2mm]
&-&g\bar{\Psi}{\gamma}^{\mu}W_{\mu}^+{\Psi}_1+
g\bar{\Psi}{\gamma}^{\mu}W_{\mu}^-{\Psi}_2\, .
\end{eqnarray}
So now the whole classical action is given by the Lagrangian
\begin{eqnarray}
{\cal L}&=&-\frac{1}{4}F_{\mu\nu}F^{\mu\nu}+\partial_{\mu}\phi^*
\partial^{\mu}\phi +i\bar{\Psi}{\gamma}^{\mu}{\partial}_{\mu}{\Psi}+
\frac{1}{2}W^{+\mu}{\Delta}_{\mu\nu}W^{-\nu}
+\frac{1}{2}W^{-\mu}{\Delta}^{\dagger}_{\mu\nu}W^{+\nu}\nonumber\\
&+&\frac{1}{2}P^+\Delta P^-+\frac{1}{2}P^-{\Delta}^{\dagger}P^+
+\frac{1}{2}W^{+\mu}\Delta_{\mu}P^-+\frac{1}{2}P^-\tilde{\Delta}_{\mu}W^{+\mu}
\nonumber\\
&+&\frac{1}{2}P^+\tilde{\Delta}_{\mu}^{\dagger}W^{-\mu}
+\frac{1}{2}W^{-\mu}\Delta_{\mu}^{\dagger}P^+ 
+\bar{\Psi}_1\Delta_F{\Psi}_1
+\bar{\Psi}_2\tilde{\Delta}_F{\Psi}_2
\nonumber \\[2mm]
&-& igP^+\bar{\Psi}\gamma_5\Psi_1+igP^-\bar{\Psi}\gamma_5\Psi_2
-ig\bar{\Psi}_1 \gamma_5\Psi P^-+ig\bar{\Psi}_2\gamma_5\Psi P^+ 
\nonumber\\[2mm]
&+& g\bar{\Psi}_2{\gamma}^{\mu}W_{\mu}^+{\Psi}-
g\bar{\Psi}_1{\gamma}^{\mu}W_{\mu}^-{\Psi}
-g\bar{\Psi}{\gamma}^{\mu}W_{\mu}^+{\Psi}_1+
g\bar{\Psi}{\gamma}^{\mu}W_{\mu}^-{\Psi}_2\nonumber\\
&+& \frac{g^2}{2}W^{+\mu}W^{-\nu}(W^+_{\mu}W^-_{\nu}-W^-_{\mu} W^+_{\nu})
-\frac{g^2}{2}(W^+_{\mu}P^--W^-_{\mu}P^+)^2 \, ,
\end{eqnarray}
where
\begin{eqnarray}
\Delta_F&=&i{\gamma}^{\mu}D_{\mu}
-\frac{g}{\sqrt{2}}(1-\gamma_5)\phi-\frac{g}{\sqrt{2}}(1+\gamma_5)\phi^*-gv,
\nonumber\\
\tilde{\Delta}_F&=&i{\gamma}^{\mu}D_{\mu}^{\dagger}
+\frac{g}{\sqrt{2}}(1-\gamma_5)\phi +\frac{g}{\sqrt{2}}(1+\gamma_5)\phi^*+gv.
\end{eqnarray}

\begin{flushleft}
{\bf 3.~Low-energy Effective Action: Calculation of the Fermionic 
Determinant in Constant Field Approximation}
\end{flushleft}

The low-energy effective action is defined as follows,
\begin{eqnarray}
\mbox{exp}\left\{i\,{\Gamma}_{\rm eff}
[A_{\mu},{\phi},{\Psi},\bar{\Psi}]\right\}
{\equiv}{\int} {\cal D} W_{\mu}^+ {\cal D} W_{\mu}^-
{\cal D}\bar{\Psi}_1{\cal D}\bar{\Psi}_2 {\cal D}{\Psi}_1
{\cal D}{\Psi}_2 {\cal D}P^+{\cal D}P^- \mbox{exp}
\left[i\, \int \, d^4 x\, {\cal L}\right]\, .
\end{eqnarray}
At tree level
\begin{eqnarray}
{\Gamma}_{\rm eff}^{(0)}=S_{\rm tree}={\int}\,d^4x\left[ -\frac{1}{4}
F_{\mu\nu}F^{\mu\nu}
+{\partial}^{\mu}{\phi}^*{\partial}_{\mu}{\phi}+
i\bar{\Psi}{\gamma}^{\mu}{\partial}_{\mu}{\Psi}\right]\,.
\end{eqnarray}
At one-loop level, the integration over the heavy modes 
will lead to the determinant of the dynamical operators, 
in practice we cannot evaluate
the determinant exactly.  We shall
employ a technique called constant field 
approximation to compute the determinant, which
was first proposed by Schwinger${\cite{sch}}$ 
and later was used in
in \cite{dmnp} and \cite{ddds} to extract 
the anomaly term in $N=2$ supersymmetric Yang-Mills
theory and the one-loop effective action of supersymmetric $CP^{N-1}$ model.
To apply this method
 we first write the classical action as following form,
\begin{eqnarray}
S&=&{\int}\,d^4x\left[ -\frac{1}{4}F_{\mu\nu}F^{\mu\nu}
+{\partial}^{\mu}{\phi}^*{\partial}_{\mu}{\phi}+
i\bar{\Psi}{\gamma}^{\mu}{\partial}_{\mu}{\Psi}\right]\nonumber\\
&+&\int d^4x \frac{1}{2}(W^{+\mu},W^{-\mu},P^+,P^-)
\left(\begin{array}{cccc}
{\Delta}_{\mu\nu} & 0 &\Delta_{\mu} & 0\\ 
0 & {\Delta}^{\dagger}_{\mu\nu} &0 &\Delta_{\mu}^{\dagger} \\
\tilde{\Delta}_{\nu}^{\dagger} & 0 &\Delta &0 \\
0 & \tilde{\Delta}_{\nu} & 0 & \Delta^{\dagger} 
\end{array}\right)\left(\begin{array}{c}W^{-\nu}\\W^{+\nu}\\P^-\\P^+
\end{array} \right)\nonumber\\
&+& {\int}d^4x\frac{1}{2}\left[(\bar{\Psi}_1,\bar{\Psi}_2,
\bar{\Psi}_1,\bar{\Psi}_2)
\left(\begin{array}{cccc} 
\Delta_F & 0 & 0 & 0\\
0 & \tilde{\Delta}_F & 0 & 0\\
0 & 0 & \Delta_F & 0 \\
0 & 0 & 0 &\tilde{\Delta}_F
 \end{array}\right)
\left(\begin{array}{c}\Psi_1\\
\Psi_2\\ \Psi_1\\
\Psi_2 \end{array}\right)\right.\nonumber\\
&+&\frac{1}{2}(\bar{\Psi}_1,\bar{\Psi}_2,\bar{\Psi}_1,\bar{\Psi}_2)
\left(\begin{array}{cccc} 
-g\gamma_{\mu}\Psi & 0 & -ig\gamma_5 \Psi & 0 \\ 
0 & \gamma_{\mu}\Psi & 0 & ig\gamma_5 \Psi  \\ 
-g\gamma_{\mu}\Psi & 0 &  -ig\gamma_5 \Psi & 0\\
0 & g\gamma_{\mu}\Psi & 0 &  ig\gamma_5 \Psi\\
 \end{array}\right)
\left(\begin{array}{c} W^{-\mu}\\W^{+\mu}\\P^-\\P^+\end{array}\right)
\nonumber\\
&+&\left.\frac{1}{2}(W^{+\mu},W^{-\mu},P^+,P^-)\left(\begin{array}{cccc} 
-g\bar{\Psi}\gamma_{\mu} & 0 & g\bar{\Psi}\gamma_{\mu} & 0  \\
0 &g\bar{\Psi}\gamma_{\mu} & 0 & g\bar{\Psi}\gamma_{\mu} \\
-ig\bar{\Psi} \gamma_5 & 0 &  -ig\bar{\Psi} \gamma_5 & 0\\
0 & ig\bar{\Psi} \gamma_5 & 0 & ig\bar{\Psi} \gamma_5
\end{array}\right)
\left(\begin{array}{c}\Psi_1\\ \Psi_2\\ \Psi_1\\ \Psi_2
\end{array}\right) \right]\nonumber\\
&+&\mbox{quartic terms of massive modes}.
\end{eqnarray}
Denoting
\begin{eqnarray}
\Phi=\left(\begin{array}{c} W^{-\mu}\\W^{+\mu}\\P^-\\P^+\end{array}\right),
~~\tilde{\Psi}=\left(\begin{array}{c}\Psi_1\\ \Psi_2\\
\Psi_1\\ \Psi_2\\  \end{array}\right),
\label{eq39}
\end{eqnarray} 
we obtain
\begin{eqnarray}
S=S_{\rm tree}+{\int}d^4x\left(\Phi^{\dagger} M_{bb} \Phi+
\bar{\tilde{\Psi}}M_{fb}\Phi +\Phi^{\dagger}M_{bf}\tilde{\Psi}
+\bar{\tilde{\Psi}}M_{ff}\tilde{\Psi}\right)
\end{eqnarray}
with that
\begin{eqnarray}
M_{bb}&=&\frac{1}{2}
\left(\begin{array}{cccc}
{\Delta}_{\mu\nu} & 0 &\Delta_{\mu} & 0\\
0 & {\Delta}^{\dagger}_{\mu\nu} &0 &\Delta_{\mu}^{\dagger} \\
\tilde{\Delta}_{\nu}^{\dagger} & 0 &\Delta &0 \\
0 & \tilde{\Delta}_{\nu} & 0 & \Delta^{\dagger}
\end{array}\right),\nonumber\\
M_{fb}&=&\frac{1}{2} \left(\begin{array}{cccc}
-g\gamma_{\mu}\Psi & 0 & -ig\gamma_5 \Psi & 0 \\
0 & g\gamma_{\mu}\Psi & 0 & ig\gamma_5 \Psi  \\
-g\gamma_{\mu}\Psi & 0 &  -ig\gamma_5 \Psi & 0\\
0 & g\gamma_{\mu}\Psi & 0 &  ig\gamma_5 \Psi\\
 \end{array}\right), \nonumber\\
M_{bf}&=&\frac{1}{2} \left(\begin{array}{cccc}
-g\bar{\Psi}\gamma_{\mu} & 0 & -g\bar{\Psi}\gamma_{\mu} & 0  \\
0 &g\bar{\Psi}\gamma_{\mu} & 0 & g\bar{\Psi}\gamma_{\mu} \\
-ig\bar{\Psi} \gamma_5 & 0 &  -ig\bar{\Psi} \gamma_5 & 0\\
0 & ig\bar{\Psi} \gamma_5 & 0 & ig\bar{\Psi} \gamma_5
\end{array}\right),\nonumber\\
M_{ff}&=&\frac{1}{2} \left(\begin{array}{cccc}
\Delta_F & 0 & 0 & 0\\
0 & \tilde{\Delta}_F & 0 & 0 \\
0 & 0 & \Delta_F & 0 \\
0 & 0 & 0 &\tilde{\Delta}_F
 \end{array}\right).
\label{eq27x}
\end{eqnarray}
Using the standard formulas
\begin{eqnarray} 
I&=&\int {\cal D}b^{\dagger}{\cal D}b{\cal D}\bar{f}{\cal D}f\,
\mbox{exp}\left[\int (dx)
\left(b^{\dagger}M_{bb}b+\bar{f}M_{fb}b+b^{\dagger}M_{bf}f
+\bar{f}M_{ff}f\right)\right] \nonumber\\[2mm]
&=&\int{\cal D}b^{\dagger}{\cal D}b{\cal D}\bar{f}{\cal D}f\,
 \mbox{exp}\left\{\int (dx) \left[b^{\dagger}(M_{bb}-M_{bf}M_{ff}^{-1}M_{fb})b
\right.\right.\nonumber\\
&+&\left.\left.(\bar{f}
+b^{\dagger}M_{bf}M_{ff}^{-1})M_{ff}(M_{ff}^{-1}M_{fb}b+f)
\right]\right\}
\nonumber\\[2mm]
&=&\det M_{ff}\det{}^{-1}(M_{bb}-M_{bf}M_{ff}^{-1}M_{fb}),
\end{eqnarray}
 and
\begin{eqnarray}
\det M=\mbox{exp}\,\mbox{Tr}\,\ln M,
\end{eqnarray}
where $b$ and $f$ represent the general 
bosonic and fermionic fields, respectively,
we obtain
\begin{eqnarray}
Z[A, \phi,\Psi,\bar{\Psi}]
&=&\mbox{exp}\left\{i\,{\Gamma}_{\rm eff}
[A_{\mu},{\phi},{\Psi},\bar{\Psi}]\right\}
{\equiv}{\int} {\cal D} W_{\mu}^+ {\cal D} W_{\mu}^-
{\cal D}\bar{\Psi}_1{\cal D}\bar{\Psi}_2 {\cal D}{\Psi}_1
{\cal D}{\Psi}_2 \mbox{exp}\left[ iS \right] \nonumber\\
&=& \mbox{exp}\left[iS_{\rm tree}\right]\det M_{ff}\det{}^{-1}(M_{bb}
-M_{bf}M_{ff}^{-1}M_{fb})\nonumber\\
&=&\mbox{exp}\left[iS_{\rm tree}+\mbox{Tr}\ln M_{ff}-
\mbox{Tr}\ln(M_{bb}-M_{bf}M_{ff}^{-1}M_{fb})\right];\nonumber\\
{\Gamma}_{\rm eff}&=&S_{\rm tree}-i\left[\mbox{Tr}\ln M_{ff}-
\mbox{Tr}\ln(M_{bb}-M_{bf}M_{ff}^{-1}M_{fb})\right].
\label{eq27} 
\end{eqnarray}
Now we evaluate above determinants. Let us first see the fermionic 
part, since $M_{ff}$ in the form of reducible matrix, 
\begin{eqnarray} 
\det M_{ff}=\frac{1}{16}(\det\Delta_F)^2(\det\tilde{\Delta}_F)^2
=\frac{1}{16}\mbox{exp}[2(\mbox{Tr}\ln\Delta_F
+\mbox{Tr}\ln\det\tilde{\Delta}_F)].
\end{eqnarray}
We use the constant field approximation to find the eigenvalues
and eigenvectors of the above operators and hence evaluate the determinant.
As in \cite{dmnp}, we choose only the third components of the electric
and magnetic fields to be the constants different from zero,
\begin{eqnarray} 
-E_3=F^{03}{\neq}0, ~~B_3=F^{12}{\neq}0,
\end{eqnarray}
and $\phi$ is the non-vanishing constant field.
Correspondingly the potential will be
\begin{eqnarray}                
A^1=-F^{12}x_2, ~~A^3=-F^{30}x_0, ~~A^0=A^2=0.
\end{eqnarray}
In order to looking for the eigenvalues of the operators, it is necessary
to rotate to Euclidean space,
\begin{eqnarray}
&& x^4=x_4=-ix^0, ~~\partial_0=\frac{\partial}{\partial x^0}
=i\frac{\partial}{\partial x^4},\nonumber\\
&& f^{34}=f_{34}=iF^{30}, ~~f^{12}=f_{12}=F^{12}.
\end{eqnarray}
We first evaluate $\det \Delta_F$,
the eigenvalue equation for $\Delta_F$ is
\begin{eqnarray}
\Delta_F\psi(x)=\left[i\gamma^{\mu}D_{\mu}-\frac{g}{\sqrt{2}}(1-\gamma_5)\phi
-\frac{g}{\sqrt{2}}(1+\gamma_5)\phi^* -gv\right]\psi(x) 
=\omega \psi_1,
\label{eq32}
\end{eqnarray}
note that $\psi$ is four-component spinor wave function.
In order to get normalizable eigenstates, we consider the system in a
box of finite size $L$ in the $x_1$ and $x_3$ directions with periodic
boundary conditions, so the eigenvector should be following form,
\begin{eqnarray}
\psi(x)=\frac{1}{L}e^{ip_1{\cdot}x_1}e^{ip_3{\cdot}x_3}{\chi}(x_2,x_4), 
\nonumber\\
p_1=\frac{2\pi l}{L}, ~~p_3=\frac{2\pi k}{L},~~k,l=\mbox{integers}.
\end{eqnarray}
To find the eigenvalues and eigenvectors, 
we write the operators and the wave function in two-component forms,
\begin{eqnarray}
\Delta_F=\left(\begin{array}{cc} -g(\sqrt{2}\phi^*+v){\bf 1} & \Delta^-\\
\Delta^+ & -g(\sqrt{2}\phi +v){\bf 1} \end{array}\right),~~~
\chi =\left(\begin{array}{c}\chi_{1}\\ \chi_{2}\end{array}\right),
\end{eqnarray}
where ${\bf 1}$ is the $2{\times}2$ identity matrix and
\begin{eqnarray}
\Delta^{\pm}=\partial_4{\pm}i\left[\sigma_1(\partial_1+igf_{12}x_2)
+\sigma_2\partial_2+\sigma_3(\partial_3+igf_{34}x_4)\right].
\end{eqnarray}
Correspondingly, the eigenvalue equation (\ref{eq32}) 
is reduced to the following set of equations,
\begin{eqnarray}
-g(\sqrt{2}\phi^*+v){\chi}_{1}+ \Delta^-{\chi}_{2}
&=&\omega {\chi}_{1},\nonumber\\
\Delta^+{\chi}_{1}-g(\sqrt{2}\phi +v){\chi}_{2}&=&\omega {\chi}_{2},
\label{eq54}
\end{eqnarray}
and now
\begin{eqnarray}
\Delta^{\pm}=\partial_4{\mp}[\sigma_1(p_1+gf_{12}x_2)
-i\sigma_2\partial_2+\sigma_3(p_3+gf_{34}x_4)].
\end{eqnarray}
A detailed calculation and discussion on the eigenvalues 
are collected in Appendix B.
We obtain two series of eigenvalues,
\begin{eqnarray}
\omega_{\pm}(m,n)
=-g\left[\frac{(\phi +\phi^*)}{\sqrt{2}} +v\right]
{\pm}\sqrt{\frac{1}{2}g^2(\phi -\phi^*)^2-2mgf_{12}-2ngf_{34}},
\end{eqnarray} 
where for $m{\geq}1$, $n{\geq}1$ both eigenvalues are doubly degenerate,
while $\omega_{\pm}(m.0)$ and $\omega_{\pm}(0,n)$ are simply degenerate, 
and for $m=n=0$, there exists only the simply degenerate eigenvalue 
$\omega_-(0,0)$.

For the eigenvalue equation 
\begin{eqnarray}
\tilde{\Delta}_F\tilde{\psi}=\left[i\gamma^{\mu}D_{\mu}^{\dagger}
+\frac{g}{\sqrt{2}}(1-\gamma_5)\phi 
+\frac{g}{\sqrt{2}}(1+\gamma_5)\phi^*+gv \right]\tilde{\psi}
=\tilde{\omega}\tilde{\psi},
\end{eqnarray}
we obtain the eigenvalues in a similar way,
\begin{eqnarray}
\tilde{\omega}_{\pm}(m,n)
=g\left[\frac{(\phi +\phi^*)}{\sqrt{2}} +v\right]
{\pm}\sqrt{\frac{g^2}{2}(\phi -\phi^* )^2-2m gf_{12}-2n gf_{34}},
\end{eqnarray}
where the degeneracies of $\tilde{\omega}_{\pm}(m,n)$, 
$\tilde{\omega}_{\pm}(m,0)$ and $\tilde{\omega}_{\pm}(0,n)$ with
$m{\geq}1$, $n{\geq}1$ is the same as those of the $\omega$s. There
 still only exists the simply degenerate eigenvalue $\omega_-(0,0)$.    

With above eigenvalues, we are able to evaluate
$\mbox{Tr}\ln\Delta_F$ and $\mbox{Tr}\ln\tilde{\Delta}_F$, in general, 
\begin{eqnarray}
\mbox{Tr}\ln\Delta_F=\ln\det\Delta_F
=\ln\left[\Pi\omega_{\pm (lk)}(m,n)\right]^r
=\sum_{l,k=-\infty}^{+\infty}\sum_{m,n=0}^{\infty} r
\ln\omega_{\pm (lk)}(m,n),
\end{eqnarray}
where $r$ is the degeneracy of $\omega_{\pm}(m,n)$.
Due to he relation $x_2=2\pi l/(gf_{12} L)$ and $x_4=2\pi k/(gf_{34} L)$,
the summation over the momentum $k$ and $l$ is actually equivalent to
the integration over $x_2$ and $x_4$, since the fields are constants,
this integration will yield only a Euclidean 
space volume factor, which tends to infinity
in the continuous limit ($L{\rightarrow}\infty$), 
\begin{eqnarray}
\sum_{l,k}=\frac{L^2}{4\pi^2}g^2f_{12}f_{34}{\int}dx_2dx_4
=\frac{V}{4\pi^2}g^2f_{12}f_{34}.
\end{eqnarray}
Consider the degeneracy of each eigenvalue, we have
\begin{eqnarray}
\mbox{Tr}\ln\Delta_F&=&\frac{V}{4\pi^2}g^2f_{12}f_{34}
\left\{\left[\ln\omega_-(0,0)+\sum_{m=1}^{\infty}\ln\omega_+(m,0)
+\sum_{n=1}^{\infty}\ln\omega_+(0,n)\right.\right.\nonumber\\
&+&\left.\left.2\sum_{m,n=1}^{\infty}\ln\omega_+(m,n)
\right] +\left[\sum_{m=1}^{\infty}\ln\omega_-(m,0)+
\sum_{n=1}^{\infty}\ln\omega_-(0,n)
+2\sum_{m,n=1}^{\infty}\ln\omega_-(m,n)
\right]\right\}\nonumber\\
&=&\frac{V}{4\pi^2}ge^2f_{12}f_{34}\left\{\ln\omega_-(0,0)
+\sum_{m=1}^{\infty}\ln[\omega_+(m,0)\omega_-(m,0)]
\right.\nonumber\\
&+&\left.\sum_{n=1}^{\infty}\ln[\omega_+(0,n)\omega_-(0,n)]
+2\sum_{m,n=1}^{\infty}\ln\left[\omega_+(m,n)\omega_-(m,n)\right]
\right\} \nonumber\\ 
&=&\frac{V}{4\pi^2}g^2f_{12}f_{34}
\left\{\ln[-g(\sqrt{2}\phi +v)]+
\sum_{m=1}^{\infty}
\ln [g^2(\sqrt{2}\phi^*+v)(\sqrt{2}\phi+v)+2mgf_{12}]\right.\nonumber\\
&+&\sum_{n=1}^{\infty}\ln [g^2(\sqrt{2}\phi^*+v)(\sqrt{2}\phi+v)
+2ngf_{34}]\nonumber\\
&+&\left.2\sum_{m,n=1}^{\infty}
\ln [g^2(\sqrt{2}\phi^*+v)(\sqrt{2}\phi+v)+2mgf_{12}+2ngf_{34}]\right\}.
\end{eqnarray}
Similarly we have 
\begin{eqnarray}
\mbox{Tr}\ln\tilde{\Delta}_F
&=&\frac{V}{4\pi^2}g^2f_{12}f_{34}\left\{\ln\left[g(\sqrt{2}\phi^*+v) \right]
+\sum_{m=1}^{\infty}
\ln [g^2(\sqrt{2}\phi^*+v)(\sqrt{2}\phi+v)+2mgf_{12}]\right.\nonumber\\
&+&\sum_{n=1}^{\infty}\ln [g^2(\sqrt{2}\phi^*+v)(\sqrt{2}\phi+v)+2ngf_{34}]
\nonumber\\
&+&\left.2\sum_{m,n=1}^{\infty}\ln [g^2(\sqrt{2}\phi^*+v)(\sqrt{2}\phi+v)
+2mgf_{12}+2ngf_{34}]\right\}.
\end{eqnarray}
Thus we finally obtain
\begin{eqnarray}
\mbox{Tr}\ln\Delta_F +\mbox{Tr}\ln\tilde{\Delta}_F
&=&\frac{V}{4\pi^2}g^2f_{12}f_{34}
\left\{\ln\left[\frac{\sqrt{2}\phi^*+v }{\sqrt{2}\phi +v }
\right]\right.\nonumber\\
&+&2\sum_{m=1}^{\infty} \ln [g^2(\sqrt{2}\phi^*+v)(\sqrt{2}\phi+v)
+2mgf_{12}]\nonumber\\
&+&2\sum_{n=1}^{\infty}\ln [g^2(\sqrt{2}\phi^*+v)(\sqrt{2}\phi+v)
+2ngf_{34}]\nonumber\\
&+& \left.4\sum_{m,n=1}^{\infty}\ln [g^2(\sqrt{2}\phi^*+v)(\sqrt{2}\phi+v)
+2mgf_{12}+2ngf_{34}]  \right\}.
\end{eqnarray}
Using the formula in the proper-time regularization,
\begin{eqnarray}
\ln \alpha=-\int^{\infty}_{{1}/{\Lambda^2}}\frac{ds}{s}e^{-\alpha s}
\end{eqnarray}
with $\Lambda^2$ is the cut-off to regularize the infinite sum,
we have
\begin{eqnarray}
&&\mbox{Tr}\ln\Delta_F +\mbox{Tr}\ln\tilde{\Delta}_F
=\frac{V}{4\pi^2}g^2f_{12}f_{34}\left\{
\ln\left[\frac{\sqrt{2}\phi^*+v}{\sqrt{2}\phi+v) }\right]
-2\int^{\infty}_{{1}/{\Lambda^2}}
\frac{ds}{s}e^{-g^2(\sqrt{2}\phi^*+v)(\sqrt{2}\phi+v) s}\right.\nonumber\\
&{\times}&\left.\left[\sum_{m=1}^{\infty}e^{-2mgf_{12}s}
+\sum_{n=1}^{\infty}e^{-2ngf_{34}s}
+2\sum_{m,n=1}^{\infty}e^{-(2mgf_{12}+2ngf_{34})s}\right]\right\}
\nonumber\\
&=&\frac{V}{4\pi^2}g^2f_{12}f_{34}
\left\{\ln\left[\frac{\sqrt{2}\phi^*+v }{\sqrt{2}\phi+v }\right]
-2\int^{\infty}_{{1}/{\Lambda^2}}
\frac{ds}{s}e^{-g^2(\sqrt{2}\phi^*+v)(\sqrt{2}\phi+v) s}
\left[\frac{e^{-gf_{34}s}}{\sinh (gf_{34}s)}\right.\right.\nonumber\\
&+&\frac{e^{-gf_{12}s}}{\sinh (gf_{12}s)}
+\left.\left.\frac{e^{-(gf_{12}+gf_{34})s}}
{\sinh (gf_{12}s)\,\sinh (gf_{34}s)}\right]\right\}
\nonumber\\
&=&\frac{V}{4\pi^2}g^2f_{12}f_{34}
\left[\ln\left(\frac{\sqrt{2}\phi^*+v}{\sqrt{2}\phi+v }\right)
-\int^{\infty}_{{1}/{\Lambda^2}}\frac{ds}{s}
e^{-g^2(\sqrt{2}\phi^*+v)(\sqrt{2}\phi+v)s}\coth(gf_{12}s)
\coth(gf_{34}s)\right]\nonumber\\
&=&\frac{V}{4\pi^2}g^2f_{12}f_{34}
\left[\ln\left(\frac{\sqrt{2}\phi^*+v}{\sqrt{2}\phi+v}\right)\right.\nonumber\\
&-&\left.\int^{\infty}_{{1}/{\Lambda^2}}\frac{ds}{s}
e^{-g^2(\sqrt{2}\phi^*+v)(\sqrt{2}\phi+v) s}
\frac{\cosh[g(f_{12}+f_{34})s]+\cosh[g(f_{12}-f_{34})s]}
{\cosh[g(f_{12}+f_{34})s]-\cosh[g(f_{12}-f_{34})s]}\right]\, ,
\end{eqnarray}
where we have used
\begin{eqnarray}
\sum_{m=1}^{\infty}e^{-2mt}=\frac{e^{-t}}{2\sinh t},~~~
\cosh (x+y)=\cosh x\cosh y{\pm}\sinh x\sinh y.
\end{eqnarray}
Rotating back to Minkowski space, we get
\begin{eqnarray}
&&\mbox{Tr}\ln\Delta_F +\mbox{Tr}\ln\tilde{\Delta}_F
=\frac{V}{4\pi^2}g^2iE_z H_z
\left[\ln\left(\frac{\sqrt{2}\phi^*+v }{\sqrt{2}\phi+v}\right)\right.
\nonumber\\
&-&\left.\int^{\infty}_{{1}/{\Lambda^2}}
\frac{ds}{s}e^{-g^2(\sqrt{2}\phi^*+v)(\sqrt{2}\phi+v) s}
\frac{ \cosh[g(H_z+iE_z)s]+\cosh[g(H_z-iE_z)s]}
{\cosh[g(H_z+iE_z)s]-\cosh[g(H_z-iE_z)s]}\right]\nonumber\\
&{\equiv}&\frac{V}{4\pi^2}g^2i{\bf E}{\cdot}{\bf H}
\left[\ln\left(\frac{\sqrt{2}\phi+v }{\sqrt{2}\phi^*+v }\right)
\nonumber\right.\\
&-&\left.\int^{\infty}_{{1}/{\Lambda^2}}\frac{ds}{s}
e^{-g^2(\sqrt{2}\phi^*+v)(\sqrt{2}\phi+v) s}
\frac{ \cosh[g{\bf X}s]+\cosh (g{\bf X}^*s)}
{\cosh (g{\bf X} s)-\cosh (g{\bf X}^*s)}\right],
\label{eq84}
\end{eqnarray}
where  
\begin{eqnarray}
{\bf X}{\equiv}{\bf H}+i{\bf E}.
\end{eqnarray}
To extract the divergence, we analyze the small-$s$ behaviour of
the integrand of (\ref{eq84}). According to the series expansion
\begin{eqnarray} 
\frac{ \cosh[g{\bf X}s]+\cosh[g {\bf X}^*s]}
{\cosh[g{\bf X} s]-\cosh[g{\bf X}^*s]}
&=&\frac{1}{({\bf X}^2-{\bf X}^{*2})}\left[\frac{4}{g^2s^2}+
\frac{2}{3}({\bf X}^2+{\bf X}^{*2})+{\cal O}(s^2)\right]\nonumber\\
&=&\frac{1}{F_{\mu\nu}\tilde{F}^{\mu\nu}}\left[\frac{1}{g^2s^2}+
\frac{2}{3}F_{\mu\nu}F^{\mu\nu}+{\cal O}(s^2)\right],
\label{eq84x}
\end{eqnarray}
where we have used 
\begin{eqnarray} 
i{\bf E}{\cdot}{\bf H}=\frac{1}{4}({\bf X}^2-{\bf X}^{*2})
=\frac{1}{4}F_{\mu\nu}\tilde{F}^{\mu\nu}, ~~
{\bf H}^2-{\bf E}^2=\frac{1}{2}({\bf X}^2+{\bf X}^{*2})
=\frac{1}{2}F_{\mu\nu}F^{\mu\nu}.
\end{eqnarray}
It can be easily seen from (\ref{eq84x}) that 
the integral in (\ref{eq84}) has a quadratic and 
a logarithmic divergence, so
the divergence term can be extracted by writing (\ref{eq84}) as following form,
\begin{eqnarray}
&&\mbox{Tr}\ln\Delta_F +\mbox{Tr}\ln\tilde{\Delta}_F
=\frac{V}{4\pi^2}\left\{\frac{1}{4}g^2F_{\mu\nu}\tilde{F}^{\mu\nu}
 \ln\left[\frac{\sqrt{2}\phi^*+v}{\sqrt{2}\phi +v}\right]\right.\nonumber\\
&-&\int^{\infty}_{{1}/{\Lambda^2}}
\left(\frac{1}{s^3}+\frac{1}{6}\frac{1}{s}g^2F_{\mu\nu}F^{\mu\nu}\right)
e^{-g^2(\sqrt{2}\phi^*+v)(\sqrt{2}\phi+v)s}\nonumber\\
&-&\int^{\infty}_0\frac{ds}{s^3}
e^{-g^2(\sqrt{2}\phi^*+v)(\sqrt{2}\phi+v) s}\left[\frac{1}{4}g^2s^2
F_{\mu\nu}\tilde{F}^{\mu\nu}\frac{ \cosh (g{\bf X}s)+\cosh (g{\bf X}^*s)}
{\cosh (g{\bf X} s)-\cosh (g{\bf X}^*s)}\right.\nonumber\\
&-&\left.\left.\frac{1}{s^3}
-\frac{1}{6}\frac{1}{s}g^2F_{\mu\nu}F^{\mu\nu}\right]\right\},
\label{eq86}
\end{eqnarray}
where the second term is the UV divergent term, so the cut-off $1/\Lambda^2$
is preserved to regularize the integral, while the last term is a finite term
and hence the cut-off has been removed.

Now we turn to the bosonic determinant. From (\ref{eq27}) we have  
\begin{eqnarray}
&&M_{ff}^{-1}=2\left(\begin{array}{cccc} 1/{\Delta}_F & 0 & 0 & 0\\
0 & 1/\tilde{\Delta}_F & 0 & 0 \\
0 & 0 & 1/{\Delta}_F & 0 \\
0 & 0 & 0 & 1/\tilde{\Delta}_F \end{array}\right), ~~
M_{bb}-M_{bf}M_{ff}^{-1}M_{fb}=\nonumber\\
&&\frac{1}{2}
\left(\begin{array}{cccc} \Delta_{\mu\nu}
-2g^2\bar{\psi}{\gamma}_{\mu}\frac{1}{\tilde{\Delta}_F}{\gamma}_{\nu}\psi & 0 &
\Delta_{\mu}-2ig^2\bar{\psi}{\gamma}_{\mu}\frac{1}{\Delta_F}{\gamma}_5\psi & 0\\
0 & {\Delta}^{\dagger}_{\mu\nu}
-2g^2\bar{\psi}{\gamma}_{\mu}\frac{1}{\tilde{\Delta}_F}{\gamma}_{\nu}\psi & 0
& {\Delta}^{\dagger}_{\mu}
-2ig^2\bar{\psi}{\gamma}_{\mu}\frac{1}{\tilde{\Delta}_F}{\gamma}_5\psi\\
\tilde{\Delta}_{\nu}^{\dagger}-2ig^2\bar{\psi}{\gamma}_5
\frac{1}{\Delta_F}{\gamma}_{\nu}\psi & 0 & \Delta+2g^2\bar{\psi}{\gamma}_5
\frac{1}{\Delta_F}{\gamma}_5\psi & 0\\
0 &  \tilde{\Delta}_{\nu}-2ig^2\bar{\psi}{\gamma}_5
\frac{1}{\Delta_F}{\gamma}_{\nu}\psi & 0 & \Delta^{\dagger}
+2g^2\bar{\psi}{\gamma}_5
\frac{1}{\Delta_F}{\gamma}_5\psi
\end{array}\right).\nonumber\\
\end{eqnarray}
In constant field approximation, $\bar{\psi}$ and $\psi$ can be regarded
Grassman numbers, so we can expand the bosonic determinant only 
to the quartic terms in  $\bar{\psi}$ and $\psi$.
Now the key problem is how to find the eigenvalues and eigenstates of
the operator matrix $M_{bb}-M_{bf}M_{ff}^{-1}M_{fb}$. If they could be worked
out, with the eigenvalues and eigenvectors of fermionic 
operator, we can use the technique 
developed in \cite{dmnp} to evaluate this determinant. Unfortunately
it is seems to us that in the constant field approximation
there is no possibility to find the eigenvalues and eigenstates of such 
a horrible operator matrix. This difficulty needs to be overcome.

Despite the fact that the bosonic part cannot be evaluated, from 
(\ref{eq27}) and (\ref{eq86}), the effective
Lagrangian associated with fermionic part has already shown
the feature of the perturbative part of the low-energy effective action.
First we believe that the quadratic divergence will be canceled owing
to the nonrenormalization theorem. As for the logarithmic divergence,
using
\begin{eqnarray}
\int^{\infty}_{{1}/{\Lambda^2}}e^{-g^2(\sqrt{2}\phi +v)(\sqrt{2}\phi^*+v) s}
{\sim}
-\ln\left[\frac{g^2(\sqrt{2}\phi +v)(\sqrt{2}\phi^*+v) }{\Lambda^2}\right],
\end{eqnarray}
we can see that in the Wilson effective action there is one term
proportional to
\begin{eqnarray}
F_{\mu\nu}F^{\mu\nu}\ln\left[
\frac{g^2(\sqrt{2}\phi^*+v)(\sqrt{2}\phi+v)}{\Lambda^2}\right],
\end{eqnarray}
thus the complete calculation should give the form (\ref{eq1}) 
of the low-energy
effective action, one can even guess this from the requirement of
supersymmetry since the constant field approximation and the 
proper-time regularization preserve the supersymmetry explicitly.  

There is a finite term proportional 
to $F\tilde{F}\ln\left[(\sqrt{2}\phi +v)/(\sqrt{2}\phi^*+v)\right]$ in
(\ref{eq86}), as that pointed out in \cite{dmnp}, this is the 
reflect of the axial $U(1)_R$ anomaly in the effective action. 
This anomaly term had played an important role
in Seiberg's nonperturbative analysis\cite{sei}.
 
\begin{flushleft}
{\bf 4.~Summary and Conclusion}
\end{flushleft}

In summary, we have tried to calculate the perturbative part of
the Seiberg-Witten low-energy effective action of 
$N=2$ supersymmetric Yang-Mills
theory based on the standard effective field theory technique.
It is well known that Seiberg-Witten effective action is the
cornerstone for all these new developments in supersymmetric
gauge theory, and that this effective action has been obtained
in a hardly understandable way, so it is worthwhile to
explore this effective action in a familiar method.
 Unfortunately we have confronted a insurmountable difficulty 
in evaluating bosonic operator in adopting constant field
approximation, which prevents us from getting the  complete result and
comparing with the form of (\ref{eq1}). However, the calculation of 
fermionic determinant has shown some features of the low-energy
effective action. This gives a partial verification of
the pure symmetry analysis for obtaining the low-energy effective action. 
The complete calculation presents an interesting problem 
for further investigation.

\vspace{8mm}

\acknowledgments
The  financial support of the Academy of Finland under the Project No. 37599
is greatly acknowledged. W.F.C thanks the World Laboratory, Switzerland 
for financial support. 
\newpage

\appendix

\section{the Low-energy Effective Action in Wess-Zumino Gauge}

\setcounter{equation}{0}

 To compare our result with that obtained  from non-perturbative analysis,
in this appendix we write the perturbative part of the Seiberg-Witten 
low-energy effective action (\ref{eq1}) in the Wess-Zumino gauge. 
In $N=1$ superfield, (\ref{eq1}) can be written as following form
\begin{eqnarray}
{\Gamma}= \frac{1}{16\pi}\mbox{Im} {\int}d^4 x\left[{\int}
d^2{\theta}{\cal F}''(\Phi)W^{\alpha}W_{\alpha}+{\int}d^2{\theta}
d^2\bar{\theta}{\Phi}^{\dagger}{\cal F}'(\Phi)\right]\, ,
\end{eqnarray}
where ${\Phi}$ is the $N=1$ chiral superfield
\begin{eqnarray} 
 {\Phi}&=&{\phi}(x)+i{\theta}{\sigma}^{\mu}
\bar{\theta}{\partial}_{\mu}{\phi}-\frac{1}{4}{\theta}^2
\bar{\theta}^2{\partial}^2{\phi}+\sqrt{2}{\theta}{\psi}
-\frac{i}{\sqrt{2}}{\theta}^2{\partial}_{\mu}{\psi}{\sigma}^{\mu}\bar{\theta}
+{\theta}^2F(x) \, ,\\[2mm]
{\cal F}(\Phi)&=& \frac{1}{2} {\tau}{\Phi}^2+
\frac{i}{2\pi}{\Phi}^2\ln\frac{{\Phi}^2}{{\Lambda}^2} \,,~\tau =
 \frac{\theta}{2\pi}+\frac{4{\pi}i}{g^2}\,
\label{pre1}
\end{eqnarray}
and
\begin{eqnarray}
{\cal F}'(\Phi)= \frac{d{\cal F}}{d{\Phi}}\, , ~{\cal F}''(\Phi)=
\frac{d^2{\cal F}}{d{\Phi}^2}\,.
\label{pre2} 
\end{eqnarray}
In Wess-Zumino gauge, the Abelian vector superfield  is 
\begin{eqnarray}
V=-{\theta}{\sigma}^{\mu}\bar{\theta}A_{\mu}+i{\theta}^2
(\bar{\theta}\bar{\lambda})-i{\bar{\theta}}^2(\theta\lambda)+
\frac{1}{2}{\theta}^2{\bar{\theta}}^2D\, ,
\label{sf1}
\end{eqnarray}
and the corresponding superfield strength is
\begin{eqnarray}
W_{\alpha}=-i{\lambda}_{\alpha}(y)+{\theta}_{\alpha}D-
i{\sigma}_{\alpha}^{\mu\nu\beta}{\theta}^{\beta}F_{\mu\nu}(y)
+{\theta}^2{\sigma}^{\mu\beta}_{\alpha}{\partial}_{\mu}
\bar{\lambda}_{\beta}(y)\,,
\end{eqnarray}
where $y^{\mu}
=x^{\mu}+i{\theta}{\sigma}^{\mu}\bar{\theta}$, ${\sigma}^{\mu\nu}
=\frac{1}{4}({\sigma}^{\mu}\bar{\sigma}^{\nu}-
{\sigma}^{\nu}\bar{\sigma}^{\mu})$ and $F_{\mu\nu}={\partial}_{\mu}A_{\nu}
-{\partial}_{\nu}A_{\mu}$.
Using 
\begin{eqnarray}
{\cal F}(\Phi)&=& {\cal F}(\phi)+{\cal F}'(\phi) \left[
i{\theta}{\sigma}^{\mu}
\bar{\theta}{\partial}_{\mu}{\phi}-\frac{1}{4}{\theta}^2
\bar{\theta}^2{\partial}^2{\phi}+\sqrt{2}{\theta}{\psi}
-\frac{i}{\sqrt{2}}{\theta}^2{\partial}_{\mu}{\psi}{\sigma}^{\mu}\bar{\theta}
+{\theta}^2F(x)\right]\nonumber\\[2mm]  
&+& \frac{1}{2}{\cal F}''(\phi) \left[
i{\theta}{\sigma}^{\mu}
\bar{\theta}{\partial}_{\mu}{\phi}-\frac{1}{4}{\theta}^2
\bar{\theta}^2{\partial}^2{\phi}+\sqrt{2}{\theta}{\psi}
-\frac{i}{\sqrt{2}}{\theta}^2{\partial}_{\mu}{\psi}{\sigma}^{\mu}\bar{\theta}
+{\theta}^2F(x)\right]\nonumber\\[2mm] 
&{\times}&\left[i{\theta}{\sigma}^{\mu}
\bar{\theta}{\partial}_{\mu}{\phi}-\frac{1}{4}{\theta}^2
\bar{\theta}^2{\partial}^2{\phi}+\sqrt{2}{\theta}{\psi}
-\frac{i}{\sqrt{2}}{\theta}^2{\partial}_{\mu}{\psi}{\sigma}^{\mu}\bar{\theta}
+{\theta}^2F(x)\right]
\end{eqnarray}
and the similar expansion for ${\cal F}'(\Phi)$,  we obtain
\begin{eqnarray}
{\Gamma}&=& \frac{1}{16 \pi}\mbox{Im} {\int}d^4x\,\left[
-{\cal F}''(\phi){\phi}{\partial}^2{\phi}-{\cal F}^{(3)}(\phi){\phi}
{\partial}^{\mu}{\phi}{\partial}_{\mu}{\phi}
+2 {\cal F}''(\phi){\partial}_{\mu}{\phi}{\partial}^{\mu}{\phi}\right.
\nonumber\\[2mm]
&-&{\partial}^2{\phi}{\cal F}'(\phi)  
+2i {\cal F}''(\phi){\partial}_{\mu}{\psi}{\sigma}^{\mu}\bar{\psi}
-2i {\cal F}''(\phi){\psi}{\sigma}^{\mu}{\partial}_{\mu}\bar{\psi}
+2i {\cal F}^{(3)}(\phi){\psi}{\sigma}^{\mu}\bar{\psi}{\partial}_{\mu}{\phi}
\nonumber\\[2mm]
&-&2 {\cal F}^{(3)}(\phi)F^{\dagger}{\psi}{\psi}+4F^{\dagger}F{\cal F}''(\phi)
+4i {\cal F}''(\phi){\lambda} {\sigma}^{\mu}{\partial}_{\mu}\bar{\lambda}-2
{\cal F}''(\phi)D^2\nonumber\\[2mm]
&+&4 {\cal F}''(\phi) (-F^{\mu\nu}F_{\mu\nu}
+iF_{\mu\nu}\tilde{F}^{\mu\nu})-2 \sqrt{2}i {\cal F}^{(3)}(\phi)
{\psi}{\lambda}D+2{\cal F}^{(3)}(\phi)({\lambda}{\lambda})F\nonumber\\[2mm]
&-&\left.{\cal F}^{(4)}(\phi)({\psi}{\psi})({\lambda}{\lambda})\right] \, .
\label{sw}
\end{eqnarray}
From (\ref{pre1}) and (\ref{pre2}), 
\begin{eqnarray}
{\cal F}'(\phi)=\left({\tau}+\frac{i}{\pi}\right){\phi}+
\frac{i}{\pi}{\phi}\ln\frac{\phi^2}{{\Lambda}^2} ~&,&~
{\cal F}''(\phi)={\tau}+\frac{3i}{\pi}+
\frac{i}{\pi}\ln\frac{\phi^2}{{\Lambda}^2}\, ,
\nonumber\\[2mm]
{\cal F}^{(3)}(\phi)=\frac{2i}{\pi}\,\frac{1}{\phi} ~&,&~
{\cal F}^{(4)}(\phi)=-\frac{2i}{\pi}\,\frac{1}{{\phi}^2}\,,
\end{eqnarray}
and rescaling the field $X{\longrightarrow}gX$ with $X=(A,{\phi},
{\lambda},{\psi})$.
we can write (\ref{sw}) as follows
\begin{eqnarray}
{\Gamma}&=& \int d^4x\left\{[-8{\pi}{\phi}{\partial}^2{\phi}+
8{\pi}{\partial}_{\mu}{\phi}{\partial}^{\mu}{\phi}+8{\pi}\,i
{\partial}_{\mu}{\psi}\bar{\sigma}^{\mu}\bar{\psi}-8{\pi}\,i{\psi}
\bar{\sigma}^{\mu}{\partial}_{\mu}\bar{\psi}+16{\pi}\,i\lambda
{\sigma}^{\mu}{\partial}_{\mu}\bar{\lambda}\right.\nonumber\\[2mm]
&-&4{\pi}F_{\mu\nu}F^{\mu\nu}]
+\frac{g^2}{\pi}\,[-4{\phi}{\partial}^2{\phi}
+4{\partial}_{\mu}{\phi}{\partial}^{\mu}{\phi}+6\,i
{\partial}_{\mu}{\psi}\bar{\sigma}^{\mu}\bar{\psi}-6{\pi}\,i{\psi}
\bar{\sigma}^{\mu}{\partial}_{\mu}\bar{\psi}\nonumber\\[2mm]
&+&4\,i {\psi}\bar{\sigma}^{\mu}\bar{\psi}{\partial}_{\mu}{\phi}
\frac{1}{\phi}+12\,i\lambda
{\sigma}^{\mu}{\partial}_{\mu}\bar{\lambda}-3\,F_{\mu\nu}F^{\mu\nu}]
+\frac{g^2}{\pi}\,
\ln\frac{\phi^2}{\Lambda^2}
[-2{\phi}{\partial}^2{\phi}
+2{\partial}_{\mu}{\phi}{\partial}^{\mu}{\phi}\nonumber\\[2mm]
&+&2\,i{\partial}_{\mu}{\psi}\bar{\sigma}^{\mu}\bar{\psi}
-2\,i{\psi}\bar{\sigma}^{\mu}{\partial}_{\mu}\bar{\psi}
+4 \,i\lambda{\sigma}^{\mu}{\partial}_{\mu}\bar{\lambda}
-F_{\mu\nu}F^{\mu\nu}]+\frac{g^2}{8{\pi}^2}\frac{1}{{\phi}^2}
(\psi\psi)(\lambda\lambda)\nonumber\\[2mm]
&-&\frac{1}{2}
\left(1+\frac{3g^2}{4{\pi}^2}+\frac{g^2}{8{\pi}^2}
\ln\frac{{\phi}^2}{{\Lambda}^2}\right)
 -\frac{\sqrt{2}g^2}{4{\pi}^2}
\frac{i(\psi\lambda)D}{\phi}-\frac{g^2}{4{\pi}^2}\frac{F^{\dagger}
(\psi\psi)}{\phi}
\nonumber\\[2mm]
&+&F^{\dagger}F\left(1+\frac{3g^2}{4{\pi}^2}
+\frac{g^2}{4{\pi}^2}\ln\frac{\phi ^2}{
\lambda ^2}\right)+\frac{g^2}{4\pi ^2}\frac{(\lambda\lambda)F}{\phi},
\label{eqa8}
\end{eqnarray}
where we set the vacuum angle ${\theta}=0$. Using the equations of motion 
$F$, $F^{\dagger}$ and $D$ derived from (\ref{eqa8})
\begin{eqnarray}
F \left(1+\frac{3g^2}{4{\pi}^2}+\frac{g^2}{8{\pi}^2}
\ln\frac{{\phi}^2}{{\Lambda}^2}\right)-\frac{g^2}{4{\pi}^2}
\frac{\psi\psi}{\phi}&=&0, \nonumber\\[2mm]
F^{\dagger}\left(1+\frac{3g^2}{4{\pi}^2}+\frac{g^2}{8{\pi}^2}
\ln\frac{{\phi}^2}{{\Lambda}^2}\right)+\frac{g^2}{4{\pi}^2}
\frac{\lambda\lambda}{\phi}&=&0,\nonumber\\[2mm]
D\left(1+\frac{3g^2}{4{\pi}^2}+\frac{g^2}{8{\pi}^2}
\ln\frac{{\phi}^2}{{\Lambda}^2}\right)+i\frac{\sqrt{2}g^2}{4{\pi}^2}
\frac{\psi\lambda}{\phi}&=&0,
\end{eqnarray}
and the algebraic manipulations
\begin{eqnarray}
{\int}d^4x\,\ln\frac{g^2{\phi}^2}{{\Lambda}^2}
{\phi}{\partial}^2{\phi}&=&
{\int}d^4x\,\left\{{\partial}_{\mu}
\left[\ln\frac{g^2{\phi}^2}{{\Lambda}^2}
{\phi}{\partial}^{\mu}{\phi}\right]
-{\partial}_{\mu}\ln\frac{g^2{\phi}^2}{{\Lambda}^2}
{\phi}{\partial}^{\mu}{\phi}-
\ln\frac{g^2{\phi}^2}{{\Lambda}^2}
{\partial}_{\mu}{\phi}{\partial}^{\mu}{\phi}\right\}\nonumber\\[2mm]
&=& -{\int}d^4x\,\left[2+ \ln\frac{g^2{\phi}^2}{{\Lambda}^2}\right]
{\partial}_{\mu}{\phi}{\partial}^{\mu}{\phi},\\
{\int}d^4x\,{\psi}\bar{\sigma}^{\mu}\bar{\psi}\frac{{\partial}_{\mu}
\phi}{\phi}&=&\frac{1}{2}
{\int}d^4x\,{\psi}\bar{\sigma}^{\mu}\bar{\psi}{\partial}_{\mu}
\ln\frac{g^2{\phi}^2}{{\Lambda}^2}\nonumber\\[2mm]
&=& -\frac{1}{2}{\int}d^4x\,
\ln\frac{g^2{\phi}^2}{{\Lambda}^2}\,[
{\partial}_{\mu}{\psi}\bar{\sigma}^{\mu}\bar{\psi}
+{\psi}\bar{\sigma}^{\mu}{\partial}_{\mu}\bar{\psi}]\, ,
\end{eqnarray}
we obtain 
\begin{eqnarray}
{\Gamma}&=&  \int d^4x\left\{\left[{\partial}_{\mu}{\phi}
{\partial}^{\mu}{\phi}+{\lambda}
{\sigma}^{\mu}{\partial}_{\mu}\bar{\lambda}+i\bar{\psi}\bar{\sigma}^{\mu}
{\partial}_{\mu}{\psi}
-\frac{1}{4}F_{\mu\nu}F^{\mu\nu}\right]\right.\nonumber\\[2mm]            
&+& \frac{3g^2}{4 {\pi}^2}\left[
{\partial}_{\mu}{\phi}{\phi}+\bar{\lambda}
{\sigma}^{\mu}{\partial}_{\mu}\bar{\lambda}+i\bar{\psi}\bar{\sigma}^{\mu}
{\partial}_{\mu}{\psi}-\frac{1}{4}F_{\mu\nu}F^{\mu\nu}\right]
\nonumber\\[2mm]
&+&\frac{g^2}{4{\pi}^2}
\ln\frac{g^2\phi^2}{{\Lambda}^2}\left[ 
{\partial}_{\mu}{\phi}{\partial}^{\mu}{\phi}+i\bar{\lambda}
{\sigma}^{\mu}{\partial}_{\mu}\bar{\lambda}+i\bar{\psi}\bar{\sigma}^{\mu}
{\partial}_{\mu}{\psi}-\frac{1}{4}
F_{\mu\nu}F^{\mu\nu}\right] 
+\left.\frac{g^2}{8{\pi}^2}\frac{(\lambda\lambda)(\psi\psi)}{\phi^2}
\right\} \,. 
\label{swe}
\end{eqnarray}         
In four-component form
$$ {\Psi}=\left(\begin{array}{l}
\psi\\ 
\bar{\lambda} \end{array} \right)\,, ~\bar{\Psi}=(\lambda , \bar{\psi})\,, ~
{\gamma}^{\mu}=\left(\begin{array}{ll}
0 & {\sigma}^{\mu}\\
\bar{\sigma}^{\mu} & 0 \end{array} \right)\, , $$
especially using the fact that for $N=2$ Abelian multiplet, $\Psi$ should
be a Majorana spinor, $\psi=\lambda$ and $\bar{\psi}=\bar{\lambda}$
 and $(\lambda\lambda)(\psi\psi)=(\psi\psi)^2=1/4(\bar{\Psi}{\Psi})^2$, thus
 (\ref{swe}) can be written as following final form 
\begin{eqnarray}
{\Gamma}&=& \int d^4x\left\{\left[1+\frac{3g^2}{4{\pi}^2}
+\frac{g^2}{4{\pi}^2}\ln\frac{\phi^2}{{\Lambda}^2}\right]\,\left[
{\partial}_{\mu}{\phi}
{\partial}^{\mu}{\phi}+
i\bar{\Psi}\bar{\gamma}^{\mu}
{\partial}_{\mu}{\Psi}-\frac{1}{4}F_{\mu\nu}F^{\mu\nu}\right]\right.
\nonumber\\[2mm]
&+&\left.\frac{g^4}{32{\pi}^2}\frac{(\bar{\Psi}{\Psi})^2}{g^2{\phi}^2}\right\}.
\label{eq:last}
\end{eqnarray}
(\ref{eq:last}) is the perturbative part of the low-energy effective 
action in Wess-Zumino gauge given by Seiberg.

\section{Calculation of the Eigenvalues of Fermionic Operator}

\setcounter{equation}{0}

In this appendix we present a detailed calculation on the eigenvalues
of fermionic operator $\Delta_F$. This can be regarded as an alternative
method to the calculation in ref.\cite{dmnp}.

First eq.(\ref{eq54}) implies that
\begin{eqnarray}
\Delta^-\Delta^+{\chi}_{1}(x_2,x_4)
&=&\left[{\omega}+g(\sqrt{2}\phi^*+v)\right]
\left[{\omega}+g(\sqrt{2}\phi +v) \right]{\chi}_{1}(x_2,x_4),
\label{eq56} \nonumber\\
\Delta^+\Delta^-{\chi}_{2}(x_2,x_4)
&=&\left[{\omega}+g(\sqrt{2}\phi^*+v)\right]
\left[{\omega}+g(\sqrt{2}\phi +v) \right] {\chi}_{2}(x_2,x_4)
\label{eq57}
\end{eqnarray}
with
\begin{eqnarray}
\Delta^+\Delta^-&=&\partial^2_2-g^2f_{12}^2(x_2+\frac{p_1}{gf_{12}})^2+
\partial^2_4-g^2f_{34}^2(x_4+\frac{p_3}{gf_{34}})^2+g\sigma_3(f_{12}+f_{34})
\nonumber\\
&=&-H_{12}-H_{34}+g\sigma_3(f_{12}+f_{34}),
\nonumber\\
\Delta^-\Delta^+ &=&\partial^2_2-g^2f_{12}^2(x_2+\frac{p_1}{gf_{12}})^2+
\partial^2_4-g^2f_{34}^2(x_4+\frac{p_1}{gf_{34}})^2+g\sigma_3(f_{12}-f_{34})
\nonumber\\
&=&-H_{12}-H_{34}+g\sigma_3(f_{12}-f_{34}),
\end{eqnarray}
where $H_{12}$ and $H_{34}$ are the Hamiltonian operators of
two independent harmonic oscillators,
\begin{eqnarray}
H_{12}&=&-\frac{\partial^2}{\partial x^2_2}
+g^2f_{12}^2(x_2+\frac{p_1}{gf_{12}})^2=-\frac{\partial^2}{\partial \xi^2_2}
+\Omega_{12}^2\xi^2_2, ~\xi_2{\equiv}x_2+\frac{p_1}{gf_{12}},
~\Omega_{12}{\equiv}|gf_{12}|;\nonumber\\
H_{34}&=&-\frac{\partial^2}{\partial x^2_4}
+g^2f_{34}^2(x_4+\frac{p_3}{gf_{34}})^2=-\frac{\partial^2}{\partial \xi^2_4}
+\Omega_{34}^2\xi^2_4, ~\xi_4=x_4+\frac{p_3}{gf_{34}},
~\Omega_{34}=|gf_{34}|.
\end{eqnarray}
Eq.(\ref{eq57}) means that the eigenvalue and the eigenvector of
$\Delta_F$ must be that of $\Delta^+\Delta^-$ and $\Delta^-\Delta^+$, while
the reverse may be not true. We can make use of the eigenvalue
and the eigenvector of $\Delta^+\Delta^-$ and $\Delta^-\Delta^+$ to find
the ones of $\Delta_F$. Like the usual method dealing with harmonic
oscillator, defining the destruction and creation operators,
\begin{eqnarray}
a_2&=&\frac{1}{\sqrt{2}}
\left(\sqrt{\Omega_{12}}\xi_2+\frac{1}{\sqrt{\Omega_{12}}}
\frac{\partial}{\partial\xi_2}\right), ~~a_2^{\dagger}
=\frac{1}{\sqrt{2}}\left(\sqrt{\Omega_{12}}\xi_2-\frac{1}{\sqrt{\Omega_{12}}}
\frac{\partial}{\partial\xi_2}\right),~~[a_2,a_2^{\dagger}]=1,\nonumber\\
a_4&=&\frac{1}{\sqrt{2}}
\left(\sqrt{\Omega_{34}}\xi_4+\frac{1}{\sqrt{\Omega_{34}}}
\frac{\partial}{\partial\xi_4}\right), ~~a_4^{\dagger}
=\frac{1}{\sqrt{2}}\left(\sqrt{\Omega_{34}}\xi_4+\frac{1}{\sqrt{\Omega_{34}}}
\frac{\partial}{\partial\xi_4}\right),~~[a_4,a_4^{\dagger}]=1,\nonumber\\
\end{eqnarray}
we have the Hamiltonian operators and their eigenstates 
in Fock space,
\begin{eqnarray}
H_{12}&=&\Omega_{12}(2a_2a_2^{\dagger}+1), ~~|n_{12}{\rangle}=
\frac{1}{\sqrt{n_{12}!}}(a_2^{\dagger})^{n_{12}}|0_{12}{\rangle},\nonumber\\ 
a_2|0_{12}{\rangle}&=&0,~~
H_{12}|n_{12}{\rangle}=\Omega_{12}(2n_{12}+1)|n_{12}{\rangle};\nonumber\\
H_{34}&=&\Omega_{34}(2a_2a_2^{\dagger}+1), ~~|n_{34}{\rangle}=
\frac{1}{\sqrt{n_{34}!}}(a_2^{\dagger})^{n_{34}}|0_{34}{\rangle},\nonumber\\ 
a_2|0_{34}{\rangle}&=&0,~~
H_{34}|n_{34}{\rangle}=\Omega_{34}(2n_{34}+1)|n_{34}{\rangle}.
\end{eqnarray}
Correspondingly, the operators $\Delta^+$ and $ \Delta^-$
can be written in terms of the destruction and creation operators,
\begin{eqnarray}
\Delta^+&=&i\sigma_2\frac{\partial}{\partial \xi_2}-gf_{12}\xi_2\sigma_1
+\frac{\partial}{\partial \xi_4}-gf_{34}\xi_4\sigma_3\nonumber\\
&=&\left(\begin{array}{cc}
\sqrt{\frac{\Omega_{34}}{2}}(a_4-a_4^{\dagger})
-\frac{gf_{34}}{\sqrt{2\Omega_{34}}}(a_4+a_4^{\dagger}) &
\sqrt{\frac{\Omega_{12}}{2}}(a_2-a_2^{\dagger})
-\frac{gf_{12}}{\sqrt{2\Omega_{12}}}(a_2+a_2^{\dagger}) \\
-\sqrt{\frac{\Omega_{12}}{2}}(a_2-a_2^{\dagger})
-\frac{gf_{12}}{\sqrt{2\Omega_{34}}}(a_2+a_2^{\dagger}) &
\sqrt{\frac{\Omega_{34}}{2}}(a_4-a_4^{\dagger})
-\frac{gf_{34}}{\sqrt{2\Omega_{34}}}(a_4+a_4^{\dagger}) \end{array}\right);
\nonumber\\
\Delta^-&=&-i\sigma_2\frac{\partial}{\partial \xi_2}+gf_{12}\xi_2\sigma_1
+\frac{\partial}{\partial \xi_4}+gf_{34}\xi_4\sigma_3\nonumber\\
&=&\left(\begin{array}{cc}
\sqrt{\frac{\Omega_{34}}{2}}(a_4-a_4^{\dagger})
+\frac{gf_{34}}{\sqrt{2\Omega_{34}}}(a_4+a_4^{\dagger}) &
-\sqrt{\frac{\Omega_{12}}{2}}(a_2-a_2^{\dagger})
+\frac{gf_{12}}{\sqrt{2\Omega_{12}}}(a_2+a_2^{\dagger}) \\
\sqrt{\frac{\Omega_{12}}{2}}(a_2-a_2^{\dagger})
+\frac{gf_{12}}{\sqrt{2\Omega_{12}}}(a_2+a_2^{\dagger}) &
\sqrt{\frac{\Omega_{34}}{2}}(a_4-a_4^{\dagger})
-\frac{gf_{34}}{\sqrt{2\Omega_{34}}}(a_4+a_4^{\dagger}) \end{array}\right).
\end{eqnarray}
Four different cases should be considered, respectively,
\begin{itemize}
\item[1.] ~$gf_{12}>0$, $gf_{34}>0$; $\Omega_{12}=gf_{12}$, 
$\Omega_{34}=gf_{34}$;
\begin{eqnarray}
\Delta^+ &=&\left(\begin{array}{cc}
-\sqrt{2\Omega_{34}}a_4^{\dagger} & -\sqrt{2\Omega_{12}}a_2^{\dagger}\\
-\sqrt{2\Omega_{12}}a_2 & \sqrt{2\Omega_{34}}a_4\end{array}\right), ~~
\Delta^- =\left(\begin{array}{cc}
\sqrt{2\Omega_{34}}a_4 & \sqrt{2\Omega_{12}}a_2^{\dagger}\\
\sqrt{2\Omega_{12}}a_2 & -\sqrt{2\Omega_{34}}a_4^{\dagger} \end{array}\right);
\end{eqnarray}
\item[2.] ~$gf_{12}>0$, $gf_{34}<0$; $\Omega_{12}=gf_{12}$, 
$\Omega_{34}=-gf_{34}$;
\begin{eqnarray}
\Delta^+ &=&\left(\begin{array}{cc}
\sqrt{2\Omega_{34}}a_4 & -\sqrt{2\Omega_{12}}a_2^{\dagger}\\
-\sqrt{2\Omega_{12}}a_2 & -\sqrt{2\Omega_{34}}a_4^{\dagger}
\end{array}\right), ~~
\Delta^- =\left(\begin{array}{cc}
-\sqrt{2\Omega_{34}}a_4^{\dagger} & \sqrt{2\Omega_{12}}a_2^{\dagger}\\
\sqrt{2\Omega_{12}}a_2 & -\sqrt{2\Omega_{34}}a_4 \end{array}\right);
\end{eqnarray}
\item[3.] ~$gf_{12}<0$, $gf_{34}>0$; $\Omega_{12}=-gf_{12}$, 
$\Omega_{34}=gf_{34}$;
\begin{eqnarray}
\Delta^+ &=&\left(\begin{array}{cc}
-\sqrt{2\Omega_{34}}a_4^{\dagger} & \sqrt{2\Omega_{12}}a_2\\
\sqrt{2\Omega_{12}}a_2^{\dagger} & -\sqrt{2\Omega_{34}}a_4
\end{array}\right), ~~
\Delta^- =\left(\begin{array}{cc}
\sqrt{2\Omega_{34}}a_4 & -\sqrt{2\Omega_{12}}a_2\\
-\sqrt{2\Omega_{12}}a_2^{\dagger} & -\sqrt{2\Omega_{34}}a_4^{\dagger} 
\end{array}\right);
\end{eqnarray}
\item[4.] ~$gf_{12}<0$, $gf_{34}>0$; $\Omega_{12}=-gf_{12}$, 
$\Omega_{34}=gf_{34}$;
\begin{eqnarray}
\Delta^+ &=&\left(\begin{array}{cc}
\sqrt{2\Omega_{34}}a_4 & \sqrt{2\Omega_{12}}a_2 \\
\sqrt{2\Omega_{12}}a_2^{\dagger} & -\sqrt{2\Omega_{34}}a_4^{\dagger}
\end{array}\right), ~~
\Delta^- =\left(\begin{array}{cc}
-\sqrt{2\Omega_{34}}a_4^{\dagger} & -\sqrt{2\Omega_{12}}a_2\\
-\sqrt{2\Omega_{12}}a_2^{\dagger} & \sqrt{2\Omega_{34}}a_4^{\dagger} 
\end{array}\right).
\end{eqnarray}
\end{itemize}
Now we look for the eigenvalues of $\Delta_F$ with aid of the ones of 
$\Delta^+\Delta^-$ and $\Delta^-\Delta^+$. Due to Eq.(\ref{eq57}) we have
\begin{eqnarray}
\Delta^-\Delta^+|{\chi}_{1}{\rangle}
&=&\left[{\omega}+g(\sqrt{2}\phi^*+v)\right]
\left[{\omega}+g(\sqrt{2}\phi +v) \right]|{\chi}_{1}{\rangle},
\nonumber\\
\Delta^+\Delta^-| {\chi}_{2}{\rangle}
&=&\left[{\omega}+g(\sqrt{2}\phi^*+v)\right]
\left[{\omega}+g(\sqrt{2}\phi +v) \right] |{\chi}_{2}{\rangle}.
\end{eqnarray}
The eigenstates ${\chi}_{1}{\rangle}$ and $|{\chi}_{2}{\rangle}$ should 
be the following form,
\begin{eqnarray}
|{\chi}_{i}{\rangle}{\sim}\left(\begin{array}{l}|k,l{\rangle}\\ 
|m,n{\rangle}\end{array}\right),~~i=1, 2,
\end{eqnarray}
where $|k,l{\rangle}{\equiv}|k{\rangle}|l{\rangle}$,
 $|m,n{\rangle}{\equiv}|m{\rangle}|n{\rangle}$, $k,m$ are the quantum numbers
of the harmonic oscillator $H_{12}$ and $l,n$ are those of $H_{34}$. We first
consider the case {\bf 1}, since that
\begin{eqnarray}
\Delta^-\Delta^+\left(\begin{array}{l}|k,l{\rangle}\\ 
|m,n{\rangle}\end{array}\right)&=&
\left(\begin{array}{c}\left[
-2k \Omega_{12}-2(l+1) \Omega_{34}\right]|k,l{\rangle} \\[1mm] 
\left[-2(m+1) \Omega_{12}-2n \Omega_{34}\right]
|m,n{\rangle} \end{array}\right),\nonumber\\
\Delta^+\Delta^-\left(\begin{array}{l}|k,l{\rangle}\\ 
|m,n{\rangle}\end{array}\right)&=&
\left(\begin{array}{c} \left(-2k\Omega_{12}-2l\Omega_{34}\right)
|k,l{\rangle}\\[1mm] 
\left[-2(m+1)\Omega_{12}-2(n+1)\Omega_{34}\right]
|m,n{\rangle}\end{array}\right),
\end{eqnarray}
the common eigenstate of $\Delta^-\Delta^+$ and $\Delta^+\Delta^-$
with eigenvalue $-2m\Omega_{12}-2n\Omega_{34}$, $m,n{\geq}1$ is
\begin{eqnarray}
\left(\begin{array}{c}|\chi_1{\rangle}\\ |\chi_2{\rangle}\end{array}\right)
=\left(\begin{array}{c}\left(\begin{array}{c}
\alpha |m,n-1{\rangle}\\ \beta|m-1,n{\rangle}\end{array}\right)\\
\left(\begin{array}{c}
\gamma |m,n{\rangle}\\ \delta|m-1,n-1{\rangle}\end{array}\right)\end{array}
\right),
\end{eqnarray}
where $\alpha$, $\beta$, $\gamma$ and $\delta$ are normalization parameters.
With this eigenstate, we come to the eigenvalue equation (\ref{eq54}) in 
Fock space,
\begin{eqnarray}
&&\left(\begin{array}{cc} -g(\sqrt{2}\phi^*+v){\bf 1} & \Delta^-\\
\Delta^+ & -g(\sqrt{2}\phi +v){\bf 1} \end{array}\right)
\left(\begin{array}{c}\left(\begin{array}{c}
\alpha |m,n-1{\rangle}\\ \beta|m-1,n{\rangle}\end{array}\right)\\
\left(\begin{array}{c}
\gamma |m,n{\rangle}\\ \delta|m-1,n-1{\rangle}\end{array}\right)\end{array}
\right)
=\omega \left(\begin{array}{c}\left(\begin{array}{c}
\alpha |m,n-1{\rangle}\\ \beta|m-1,n{\rangle}\end{array}\right)\\
\left(\begin{array}{c}
\gamma |m,n{\rangle}\\ \delta|m-1,n-1{\rangle}\end{array}\right)\end{array}
\right)\nonumber\\
&&=\left(\begin{array}{c}\left[-g(\sqrt{2}\phi^*+v)\alpha +\sqrt{2n\Omega_{34}}
\gamma+\sqrt{2m\Omega_{12}}\delta \right]|m,n-1{\rangle}\\[1mm]
\left[-g(\sqrt{2}\phi^*+v)\beta +\sqrt{2m\Omega_{12}}
\gamma-\sqrt{2n\Omega_{34}}\delta \right]|m-1,n{\rangle}\\[1mm]
\left[-\sqrt{2n\Omega_{34}}\alpha-\sqrt{2m\Omega_{12}}\beta 
-g(\sqrt{2}\phi+v)\gamma \right]|m,n{\rangle}\\[1mm]
\left[-\sqrt{2m\Omega_{12}}\alpha+\sqrt{2n\Omega_{34}}\beta 
-g(\sqrt{2}\phi+v)\delta \right]|m-1,n-1{\rangle}\end{array}\right),
\label{eqb15}
\end{eqnarray}
this is in fact changed into the ordinary matrix eigenvalue problem,
\begin{eqnarray}
\left(\begin{array}{cccc} -g(\sqrt{2}\phi^*+v) & 0 & \sqrt{2n\Omega_{34}}
 & \sqrt{2m\Omega_{12}} \\[1mm]
 0 &-g(\sqrt{2}\phi^*+v)  & \sqrt{2m\Omega_{12}}
 & -\sqrt{2n\Omega_{34}} \\[1mm]
-\sqrt{2n\Omega_{34}} & -\sqrt{2m\Omega_{12}} & -g(\sqrt{2}\phi +v) & 0 \\[1mm]
-\sqrt{2m\Omega_{12}} & \sqrt{2n\Omega_{34}} & 0 & -g(\sqrt{2}\phi +v) 
\end{array}\right)
\left(\begin{array}{c}\alpha\\ \beta\\ \gamma\\ \delta\end{array}\right)
=\omega \left(\begin{array}{c}\alpha\\ \beta\\ \gamma \\ 
\delta\end{array}\right).
\label{eqb20}
\end{eqnarray}
Using the fact 
\begin{eqnarray}
\det\left(\begin{array}{cccc} K & 0 & A & B \\[1mm]
                              0 & K & B & -A \\[1mm]
                              -A & -B & L & 0 \\[1mm]
                              -B & A & 0 & L
                                \end{array}\right)
=(A^2+B^2+KL)^2,
\label{eqb21}
\end{eqnarray}
we can see that the eigenvalue $\omega$ is determined by the following equation,
\begin{eqnarray}
&& \left[\omega +g(\sqrt{2}\phi^*+v)\right]
\left[\omega +g(\sqrt{2}\phi+v)\right]+2m \Omega_{12}
+2n \Omega_{34}=0,\nonumber\\[2mm]
&&\omega_{\pm}(m,n)=-g\left[\frac{\phi+\phi^*}{\sqrt{2}}+v\right]\pm
\sqrt{\frac{1}{2}g^2(\phi-\phi^*)^2-2m \Omega_{12} -2n\Omega_{34}}. 
\end{eqnarray}
Eqs.(\ref{eqb20}) and  (\ref{eqb21}) explicitly show that $\omega_{+}(m,n)$
and  $\omega_{-}(m,n)$ with $m,n{\geq}1$ are doubly degenerate,
since for a $4{\times}4$ matrix there should exist four eigenvalues. Special 
attention should be paid to the cases when $m=0$ or $n=0$ as well as both
of them equal to zero, when we will see that the degeneracies of the eigenvalue 
will change:
\begin{itemize}  
\item $m{\geq}1$, $n=0$:~ in this case the eigenvalue equation (\ref{eqb15})
will be reduced to the following form,
\begin{eqnarray}
&&\left(\begin{array}{c} 0 \\[1mm]
\left[-g(\sqrt{2}\phi^*+v)\beta +\sqrt{2m\Omega_{12}}\gamma\right]
|m-1,0{\rangle}\\[1mm]
\left[-\sqrt{2m\Omega_{12}}\beta-g(\sqrt{2}\phi +v)\gamma\right]
|m,0{\rangle}\\[1mm] 0\end{array}\right)
=\omega \left(\begin{array}{c} 0 \\ \beta |m-1,0{\rangle} \\ 
\gamma |m,0{\rangle} \\ 0 \end{array}\right),\nonumber\\
&& \omega_{\pm}(m,0)=-g\left[\frac{\phi+\phi^*}{\sqrt{2}}+v\right]\pm
\sqrt{\frac{1}{2}g^2(\phi-\phi^*)^2-2m \Omega_{12}}.
\end{eqnarray}
The eigenvalues $\omega_{\pm}(m,0)$ are obviously simply degenerate.
\item $m=0$, $n{\geq}1$:~ in this case we have the eigenvalue 
equation as follows,
\begin{eqnarray}
&&\left(\begin{array}{c} 
\left[-g(\sqrt{2}\phi^*+v)\alpha +\sqrt{2n\Omega_{34}}\gamma\right]
|0,n-1{\rangle}\\[1mm] 0\\[1mm]
\left[-\sqrt{2n\Omega_{34}}\alpha-g(\sqrt{2}\phi +v)\gamma\right]
|0,n{\rangle}\\[1mm] 0\end{array}\right)
=\omega \left(\begin{array}{c}  \alpha |0,n-1{\rangle} \\ 0\\ 
\gamma |0,n{\rangle} \\ 0 \end{array}\right),\nonumber\\
&&\omega_{\pm}(0,n)=-g\left[\frac{\phi+\phi^*}{\sqrt{2}}+v\right]\pm
\sqrt{\frac{1}{2}g^2(\phi-\phi^*)^2-2n \Omega_{34}}.
\end{eqnarray} 
The eigenvalues $\omega_{\pm}(0,n)$ are also simply degenerate. 
\item $m=n=0$:~ the eigenvalue equation becomes very simple,
\begin{eqnarray}
\left(\begin{array}{c} 0 \\ 0\\ -g(\sqrt{2}\phi +v) |0,0{\rangle} \\ 0
\end{array}\right)&=&\omega 
\left(\begin{array}{c} 0 \\ 0\\ |0,0{\rangle} \\ 0
\end{array}\right),\nonumber\\
\omega(0,0) = -g(\sqrt{2}\phi +v)=\omega_-(0,0).
\end{eqnarray}
Thus there exists only one $\omega_-(0,0)$ and it is simply degenerate.
\end{itemize}

For the case 2, the common eigenstate of $\Delta^+\Delta^-$ and 
$\Delta^-\Delta^+$ with eigenvalue $-2m \Omega_{12}-2n \Omega_{34}$ is
\begin{eqnarray}
\left(\begin{array}{c} |\chi_1{\rangle}\\|\chi_2{\rangle}\end{array}\right)
=\left(\begin{array}{c}\left(\begin{array}{c}
\alpha |m,n{\rangle}\\ \beta|m-1,n-1{\rangle}\end{array}\right)\\
\left(\begin{array}{c}
\gamma |m,n-1{\rangle}\\ \delta|m-1,n{\rangle}\end{array}\right)\end{array}
\right).
\end{eqnarray} 
In a similar way, one can see that the eigenvalues $\omega_{\pm}(m,n)$, 
$\omega_{\pm}(m,0)$, $\omega_{\pm}(0,n)$ with $m,n{\geq}1$ and their 
degeneracies are the same as the case 1; only $\omega (0,0)$ is different,
\begin{eqnarray}
\omega (0,0)=-g(\sqrt{2}\phi^* +v)=\omega_+(0,0).
\end{eqnarray} 

As for the cases 3 and 4, the common eigenstates of $\Delta^+\Delta^-$ and 
$\Delta^+\Delta^-$ with eigenvalue $-2m \Omega_{12}-2n \Omega_{34}$ are,
respectively,
\begin{eqnarray}
&3.& ~~
\left(\begin{array}{c} |\chi_1{\rangle}\\|\chi_2{\rangle}\end{array}\right)
=\left(\begin{array}{c}\left(\begin{array}{c}
\alpha |m-1,n-1{\rangle}\\ \beta|m,n{\rangle}\end{array}\right)\\
\left(\begin{array}{c}
\gamma |m-1,n{\rangle}\\ \delta|m,n-1{\rangle}\end{array}\right)\end{array}
\right);\nonumber\\
&4.& ~~ 
\left(\begin{array}{c} |\chi_1{\rangle}\\|\chi_2{\rangle}\end{array}\right)
=\left(\begin{array}{c}\left(\begin{array}{c}
\alpha |m-1,n{\rangle}\\ \beta|m,n-1{\rangle}\end{array}\right)\\
\left(\begin{array}{c}
\gamma |m-1,n-1{\rangle}\\ \delta|m,n{\rangle}\end{array}\right)\end{array}
\right).
\end{eqnarray} 
The eigenvalues $\omega_{\pm}(m,n)$, 
$\omega_{\pm}(m,0)$, $\omega_{\pm}(0,n)$ with $m,n{\geq}1$ and their 
degeneracies are the same as the cases 1, 2, but $\omega (0,0)$'s are,
respectively,
\begin{eqnarray}
&3.& ~~\omega (0,0)=-g(\sqrt{2}\phi^* +v)=\omega_+(0,0);\nonumber\\
&4.& ~~\omega (0,0)=-g(\sqrt{2}\phi +v)=\omega_-(0,0).
\end{eqnarray} 

 The eigenvalues of $\tilde{\Delta}_F$ can be determined in a similar way;
the only difference is $g{\longrightarrow}-g$. 

It should be emphasized that these four cases are not equivalent,
since the eigenstates are different with each other. However, they give  
the identical $\det\Delta_F \det\tilde{\Delta}_F$.

\end{document}